\documentclass{aa}  

\usepackage{graphicx}
\usepackage{txfonts}
\usepackage{lipsum}
\usepackage{subcaption}         % necessary for continued figures, example in section 3
\usepackage{lscape}             % to rotate a single page table, example in appendix.
\usepackage{placeins}           % useful with \FloatBarrier, to keep 
\newcommand{\ts}{\textstyle}
\newcommand{\dyne}{{\sc Dynesty}}

\begin{document}
   \title{Technique-agnostic exoplanet demography for the \textit{Roman} era -- I. Testing a demography retrieval framework using simulated \textit{Kepler}-like transit datasets}
   \titlerunning{Technique-agnostic exoplanet demography}

   %\subtitle{Subtitle}

%%%%%%%%%%%%%%%%%%%%%%%%%%%%%%%%%%%%%%%%
% Please do not include ORCIDs next to author names.
% Only ORCIDs authenticated by individual authors in EDP Sciences editorial system will be taken into account.
% ORCIDs included here will be removed.
%%%%%%%%%%%%%%%%%%%%%%%%%%%%%%%%%%%%%%%%

\author{
Akshay~Priyadarshi\inst{\ref{in1}} \and 
Eamonn~Kerins\inst{\ref{in1}}\fnmsep\thanks{eamonn.kerins@manchester.ac.uk} \and
(RGES-PIT)\thanks{RGES-PIT: Roman Galactic Exoplanet Survey - Project Infrastructure Team}: 
Michael~D.~Albrow\inst{\ref{in1a}} \and
Jay~Anderson\inst{\ref{in2}} \and 
Etienne~Bachelet \inst{\ref{in2a},\ref{in3}} \and 
Chas~Beichman\inst{\ref{in3}} \and 
David~P.~Bennett\inst{\ref{in4},\ref{in4a}} \and 
Aparna~Bhattacharya\inst{\ref{in4},\ref{in4a}} \and 
Valerio~Bozza\inst{\ref{in5},\ref{in5a}} \and 
Chris~Brandon\inst{\ref{in6}} \and 
Sebastiano~Calchi~Novati\inst{\ref{in3}} \and 
Kylee~Carden\inst{\ref{in6a}} \and 
Sean~Carey\inst{\ref{in3}} \and 
Jessie~Christiansen\inst{\ref{in3}} \and 
Ali~Crisp\inst{\ref{in6}} \and 
William~DeRocco\inst{\ref{in4},\ref{in6a}} \and 
Scott~Gaudi\inst{\ref{in6}} \and 
Jon~Hulberg\inst{\ref{in4a},\ref{in7},\ref{in7a}} \and 
Macy~J.~Huston\inst{\ref{in8}} \and 
Stela~Ishitani Silva\inst{\ref{in9}} \and 
Somayeh~Khakpash\inst{\ref{in10}} \and 
Katarzyna~Kruszyńska\inst{\ref{in11}} \and 
Amber~Malpas\inst{\ref{in6}} \and 
Arjun~Murlidhar\inst{\ref{in6}} \and 
Casey~Lam\inst{\ref{in12}} \and 
Xavier~Lesley-Salda\~na\inst{\ref{in6}} \and
Jessica~R.~Lu\inst{\ref{in8}} \and 
Greg~Olmschenk\inst{\ref{in4},\ref{in4a}} \and 
Matthew~Penny\inst{\ref{in13}} \and 
Keivan~G.~Stassun\inst{\ref{in14}} \and 
Alexander~P.~Stephan\inst{\ref{in14}} \and 
Rachel~A.~Street\inst{\ref{in11}} \and 
Takahiro~Sumi\inst{\ref{in15}} \and 
Sean~K.~Terry\inst{\ref{in4},\ref{in4a}} \and 
Himanshu~Verma\inst{\ref{in13}} \and 
Weicheng~Zang\inst{\ref{in16}} \and 
Farzaneh~Zohrabi\inst{\ref{in13}} \and
(TRExS)\thanks{TRExS: Transits in the Roman galactic Exoplanet Survey}: Alison~Duck\inst{\ref{in6}} \and
Nestor~Espinoza\inst{\ref{in2}} \and
Kelsey~Hoffman\inst{\ref{in18}} \and
Jorge~Martinez-Palomera\inst{\ref{in17}} \and
Susan~Mullally\inst{\ref{in2}} \and
Elisa~Quintana\inst{\ref{in9}} \and
Robert~Wilson\inst{\ref{in9}}
}

%    \authorrunning{A. Priyadarshi et al.}
    
\institute{
Jodrell Bank Centre for Astrophysics, University of Manchester, Oxford Road, Manchester, M13 9PL, UK \label{in1}
\and
School of Physical and Chemical Sciences, University of Canterbury, Christchurch, New Zealand\label{in1a}
\and
Space Telescope Science Institute, 3700 San Martin Drive, Baltimore, MD 21218, USA \label{in2}
\and
Universit\'{e} Marie et Louis Pasteur, CNRS, Institut UTINAM UMR 6213, Besan\c{c}on, France \label{in2a}
\and
IPAC, Caltech, 1200 E. California Blvd., Pasadena, CA 91125, USA \label{in3}
\and
Department of Astronomy, University of Maryland, College Park, MD 20742, USA \label{in4}
\and
Code 667, NASA Goddard Space Flight Center, Greenbelt, MD 20771, USA \label{in4a}
\and
Dipartimento di Fisica ``E.R. Caianiello'', Universit\`{a} di Salerno, Via Giovanni Paolo 132, Fisciano, I-84084, Italy \label{in5}
\and
Istituto Nazionale di Fisica Nucleare, Sezione di Napoli, Via Cintia, Napoli, I-80126, Italy \label{in5a}
\and
Department of Astronomy, The Ohio State University, Columbus, OH 43210, USA \label{in6}
\and
Department of Physics \& Astronomy, The Johns Hopkins University, 3400 N. Charles Street, Baltimore, MD 21218, USA \label{in6a}
\and
SETI Institute, 189 Bernardo Ave, Suite 200, Mountain View, CA 94043, USA \label{in18}
\and
Department of Physics, Catholic University of America, Washington, DC 20064, USA \label{in7}
\and
Center for Research and Exploration in Space Science and Technology, NASA/GSFC, Greenbelt, MD 20771, USA \label{in7a}
\and
Department of Astronomy, University of California Berkeley, Berkeley, CA 94720, USA \label{in8}
\and
NASA Goddard Space Flight Center, Greenbelt, MD 20771, USA \label{in9}
\and
University of Maryland, Baltimore County, 1000 Hilltop Circle, Baltimore, Maryland, USA \label{in17}
\and
Department of Physics, Lehigh University, 16 Memorial Drive East, Bethlehem, PA 18015, USA \label{in10}
\and
Las Cumbres Observatory, 6740 Cortona Drive, Suite 102, Goleta, CA 93117, USA \label{in11}
\and
Observatories of the Carnegie Institution for Science, Pasadena, CA 91101, USA \label{in12}
\and
Department of Physics and Astronomy, Louisiana State University,
Baton Rouge, LA 70803, USA \label{in13}
\and
Department of Physics and Astronomy, Vanderbilt University, Nashville, TN 37235, USA \label{in14}
\and
Department of Earth and Space Science, Graduate School of Science, Osaka University, Osaka, 560-0043, Japan \label{in15}
\and
Center for Astrophysics, Harvard \& Smithsonian, Cambridge, MA 02138, USA \label{in16}
}

   \date{}%Received April 17, 2026}

% \abstract{}{}{}{}{}
% 5 {} token are mandatory
 
  \abstract
 {The Nancy Grace Roman Space Telescope (\textit{Roman}) will unveil for the first time the full architecture of planetary systems across Galactic distances through the discovery of up to 200,000 cool and hot exoplanets using microlensing and transit detection methods. \textit{Roman}'s huge exoplanet haul, and Galactic reach, will require new methods to leverage the full exoplanet demographic content of the combined microlensing and transit samples, given the different sensitivity bias of the techniques to planet and host properties and Galactic location.}
   {We present a framework for {\em technique-agnostic exoplanet demography}\/ (TAED) that can allow large, multi-technique exoplanet samples distributed over Galactic distance scales to be combined for demographic studies.}
   {Our TAED forward modelling and retrieval framework uses parameterised model exoplanet demographic distributions to embed planetary systems within a stellar population synthesis model of the Galaxy, enabling internally consistent forecasts to be made for all detection methods that are based on spatio-kinematic system properties. In this paper, as a first test of the TAED framework, we apply it to simulated transit datasets based on the \textit{Kepler} Data Release 25 to assess parameter recovery accuracy and method scalability for a single large homogeneous dataset.} 
   {We find that optimisation using differential evolution provides a computationally scalable framework that gives a good balance between computational efficiency and accuracy of parameter recovery.}
  % conclusions heading (optional)
   {}

   \keywords{Planets and satellites: general --
                Methods: statistical --
                Methods: data analysis
               \vspace{-4em}}
   \maketitle
   \nolinenumbers
%%%%%%%%%%%%%%%%%%%%%%%%%%%%%%%%%%%%%%%%%%%%%%%%%%%%%%%%%%%%%%

\section{Introduction}

Whilst there are now more than 6,000 confirmed exoplanets, these planets still reflect a very limited region of the exoplanet discovery space. The vast majority of known systems lie within 1~kpc of the Sun and therefore occupy only a small fraction of the Galactic stellar volume. This volume contains an even smaller fraction of all Galactic stars, and these stars are a biased subset of all Galactic stars in terms of age, chemistry and kinematics \citep[e.g.][]{Gaudi2021TheExoplanets}. The deep connection between planets and their host stars means that biases in the host star sample likely translate into biases in the characteristics of the exoplanet sample, with respect to the Galactic population as a whole. Such bias can lead to an incomplete and skewed measure of exoplanet occurrence, or a misrepresentation of the prevalent architecture and characteristics of planetary systems across the Galaxy.

The potential for bias in the stellar sample is compounded further by the use of different detection techniques that have different sensitivities to planet and host properties. The transit and radial velocity methods have so far provided us with the bulk of confirmed planetary systems, though they are samples that favour hotter (shorter period) and larger (more massive) exoplanets orbiting seismically quiet main sequence stars. Other survey techniques that are contributing significantly to current statistics include gravitational microlensing and direct imaging surveys, and it is expected that large samples of astrometrically-detected planets will soon be delivered from reductions of \textit{Gaia} observations \citep{Perryman_2014}. These techniques allow us to explore the occurrence of cooler (longer period) exoplanets, with microlensing probing cool planets over wide mass and distance scales, and direct imaging and astrometry typically targeting mostly nearby and larger planetary mass scales. 

A picture of exoplanet demographics that is reflective of the Galactic planetary population as a whole is needed to provide the strongest constraints on the physics of planet formation. We require exoplanet surveys over larger Galactic volumes that contain a representative mix of stars. We need to ensure that these representative volumes can be reached by multiple detection methods that can span the full range of exoplanet architecture, including large and small planets, and hot and cold planets. Lastly, we need methods that can tie together the full set of exoplanet demographic information provided by these different techniques, in a manner that takes full account of their variation in sensitivity to planetary orbit and bulk properties, host characteristics, and Galactic location.

The Nancy Grace Roman Space Telescope (hereafter \textit{Roman}), scheduled for launch in late 2026, will take an enormous stride in improving our understanding of exoplanet demography over Galactic distance scales. The core science aim of \textit{Roman}'s Galactic Bulge Time Domain Survey (GBTDS) is to detect and measure the masses of cool exoplanets using the gravitational microlensing method, including free-floating planets unbound from host stars \citep{Penny2019,Johnson_2020}. Simulations indicate that \textit{Roman} will be capable of finding around 1,500 planets using microlensing, providing a cool planet sample that is comparable in size to that of the \textit{Kepler} transit sample.

The same survey is also predicted to find 60,000 - 200,000 transiting planets \citep{Montet_2017, Wilson_2023} over Galactic distance scales. The GBTDS will therefore be the first exoplanet survey sensitive to essentially the full range of planetary orbits across Galactic distances for planets of Earth size and larger.  

The remaining hurdle is to employ demographic modelling methods that can be used on large exoplanet samples obtained with different detection methods probing different host star populations. These methods will need to be able to account for correlations not just between planet and host properties, but also between host properties and Galactic location. In this paper, we refer to demographic modelling that is agnostic to spatio-kinematic based detection methods (e.g. transit, radial velocity, microlensing and astrometry) as {\em technique agnostic exoplanet demography} (hereafter TAED). A TAED approach requires the integration of models of Galactic structure, stellar populations and exoplanet demography. It's an approach capable of taking full advantage of the \textit{Roman} exoplanet samples for demographic studies.

Several studies have attempted to model exoplanet demographics, incorporating the effect of detection bias. For instance, \cite{Hsu2018ImprovingComputation} use a hierarchical Bayesian modelling framework applied to their \texttt{SysSim} forward model applied to \textit{Kepler}. This is an example of a retrieval framework that is tailored to a specific survey dataset. \texttt{EPOS} \citep{Mulders2018TheSystems} is another model which aims to constrain the properties of exoplanet populations, like occurrence rates and orbital architectures, using \textit{Kepler} data. Similar to the \texttt{SysSim} model, \texttt{EPOS} samples the planets from a distribution over grids of radius and period. Both of these models could be adapted and applied to other surveys and detection techniques, but both lack a Galactic model to account for variations in host properties with Galactic location. This makes it difficult to use such models to combine results from methods with very different distance sensitivity, such as transits and microlensing.

In this paper, we develop and present a working and scalable TAED-based method for the retrieval of exoplanet demographic parameters from large exoplanet samples that may span a range of Galactic distances. In this paper, we restrict the application of our TAED framework only to relatively local samples of transiting planets simulated from a realistic \textit{Kepler}-like survey with known detection efficiency, in order to establish the efficacy and feasibility of the framework. Application to other methods (or multiple methods), and to samples that simulate \textit{Roman}'s specific detection characteristics, will be the focus of future papers. 

The present paper is structured as follows. In Section~\ref{sec:taed} we introduce our TAED approach and simulate a stellar sample for a \textit{Kepler}-like transit survey. The simulated transit sample itself is discussed in Section~\ref{sec:kepler}. The initial set of toy planetary models and the likelihood function used for retrieval are discussed in Section~\ref{sec:retrieval}. The ability of the TAED framework to successfully retrieve accurate demographic information is examined in Section~\ref{sec:retrieve-algo}, where we explore different retrieval schemes and show that those based on differential evolution appear promising. We recover optimal parameters for simple planet demographic models applied to the true \textit{Kepler} dataset in Section~\ref{sec:kepler-params}, highlighting how the TAED framework can measure demographic model parameters, but also how it shows up deficiencies in the flexibility of simple demographic models.  We summarise our findings in Section~\ref{sec:conc}.

\section{Technique-agnostic exoplanet demography} \label{sec:taed}

Using samples of planets for demographic studies that have been obtained through different detection methods requires the ability to account for differences in selection bias between methods. For methods like transit and microlensing, these differences are substantial. 

The majority of transit events to date are located within 1~kpc from us and involve planets on relatively short orbital periods. Transit samples are also biased in favour of larger planets, which yield deeper transits. Additionally, there is a strong correlation between host type and distance from the observer, with transits involving lower-mass hosts tending to be close by, as these hosts are intrinsically fainter. 

By contrast, as it does not rely on the need to detect the host star, microlensing is sensitive to planetary systems over a large range of distances, and so typically reside several kpc from the Sun. Microlensing planets are observable when their projected separation is comparable to the Einstein radius of their host, which means they are typically on much wider orbits than transiting planets. Unlike transiting planets, microlensing observations yield the planet--host mass ratio and sky-projected separation, rather than the planet--host size ratio and orbital period. Planet samples detected through microlensing are biased towards the most common types of star (low mass M and K dwarfs), whereas transit surveys have tended to target FGK-type hosts that exhibit low-levels of intrinsic variability.

Demographic studies to date that have aimed to synthesise transit and microlensing samples have usually done so by restricting their scope to sub-samples that have properties in common, such as transit and microlensing samples that involve hosts of a similar type. Whilst this approach is a valid way to proceed, the relatively small intersection in common properties between microlensing and transit samples means that many valid planets must be excluded from the joint analysis. This, in turn, strongly restricts the scope of demographic information that can be recovered.

In order to fully exploit exoplanet samples from different detection techniques, we must consider not just the demographics of planets, but also the demographics of their hosts. By considering both jointly, we can develop a technique agnostic exoplanet demographic (TAED) framework that can provide consistent forecasts for planet yields across different surveys, including surveys employing different detection techniques. In this way, a retrieval framework can then use the union, rather than the intersection, of the joint sample to recover the underlying planet demography.

The TAED approach shares some similarity to the Exoplanet Population Observation Simulator \citep[EPOS,][]{Mulders2018TheSystems} in that it employs a parameterised forward model to describe the underlying planet population. The main difference between TAED and EPOS is that in the TAED approach, we are layering our planet model on top of a model for the Galactic stellar population (i.e. the host stars). Without this additional layer, it is not possible to make meaningful joint forecasts for methods like transits and microlensing that have very different sensitivity to host type and Galactic location. 

\subsection{Host star model} \label{sec:gal-model}

We use the Besan\c{c}on Galactic Model \citep[][hereafter BGM]{Robin2003AWay,Marshall2006ModellingDimensions,2012AA...538A.106R} to describe the host star population. The BGM uses input models for the spatial distribution, masses and kinematics of stars as a function of age, together with stellar atmosphere models, to predict the observable properties of a synthetic catalogue of stars that would be seen along any given direction by magnitude and/or kinematics-limited surveys. The BGM incorporates a self-consistent 3D dust map so that stellar apparent magnitudes can be generated in a range of standard optical and infrared passbands.

\begin{figure}
	\includegraphics[width=\columnwidth]{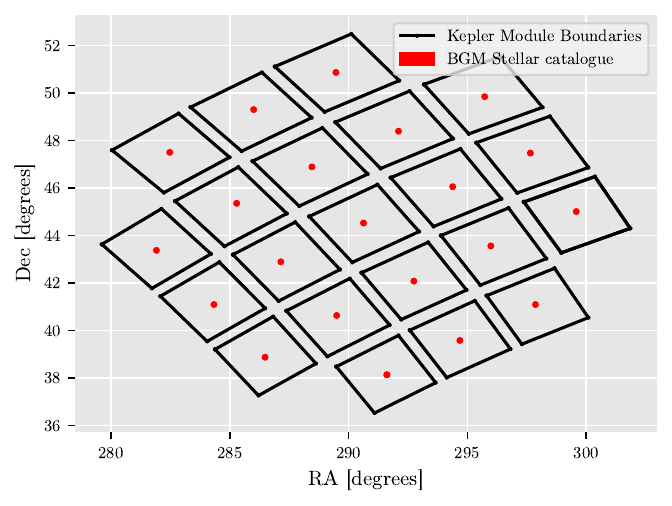}
    \caption{The \textit{Kepler} focal plane array showing the arrangement of the 21 detector modules (black outlines). The red dots show the locations used for our BGM synthetic stellar catalogues.}
    \label{fig:kepfov}
\end{figure}

For our test of the TAED framework, we simulate a \textit{Kepler}-like transit-observables by generating BGM stellar catalogues towards the \textit{Kepler} field location\footnote{\url{https://keplergo.github.io/KeplerScienceWebsite}}. We simulate stars within 0.1 square degree pencil beams centred around each of the 21 \textit{Kepler} detector array modules as shown in Figure \ref{fig:kepfov}.  This spatial sampling enables realistic simulation of host star populations across the \textit{Kepler} field, accounting for Galactic structure and stellar diversity.

\begin{figure}
	\includegraphics[width=\columnwidth]{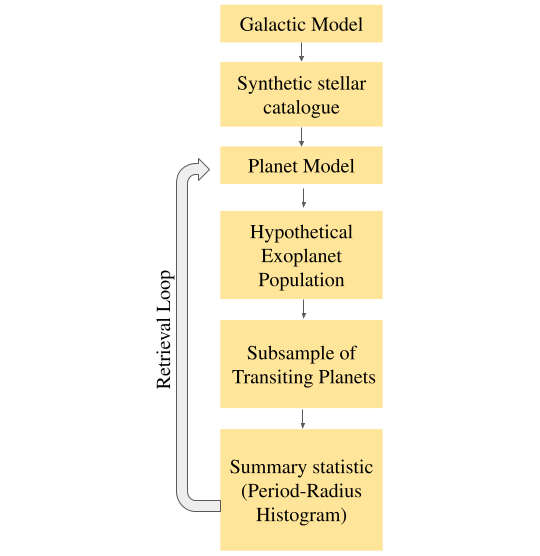}
    \caption{Schematic overview of the TAED forward-modelling and retrieval workflow. A Galactic model and parameterised planet model are forward-modelled to a transit subsample and a $(P,R_p)$ histogram, which closes the retrieval loop by iteratively updating the parameters in the planet model.}
\label{fig:taed_flow}
\end{figure}

Whilst the expected \textit{Roman} transit sample is expected to dwarf that from \textit{Kepler}, \textit{Kepler} nonetheless provides a useful testbed for simulating transit samples of significant size. Additionally, the \textit{Kepler} completeness is well characterised and is publicly available through KeplerPORTs \citep{Burke2015TERRESTRIALSAMPLE}. KeplerPORTs provides per-target detection contours giving the recovery fraction as a function of orbital period and planet radius. These contours are constructed via a multiple-event statistic (MES)-based detection-efficiency model in which the MES estimate depends on stellar parameters and per-target noise/coverage products \citep{2017ksci.rept...19B}. This allows us to simulate a transit sample using a realistic efficiency framework that we may expect to share some qualitative similarity with the framework that will be ultimately developed for \textit{Roman}, even though, quantitatively, the \textit{Roman} and \textit{Kepler} detection efficiencies will be very different. 

The output from BGM comprises catalogues of artificial stars with distance, magnitudes and colours in several Johnson-Cousins passbands, stellar effective temperature, surface gravity, mass, and radius. Colour-magnitude relations from \citet{Jordi2006EmpiricalSystems} and \citet{Brown2011KEPLERCLASSIFICATION} are used to transform from Johnson-Cousins bands to \textit{Kepler} $K_p$ magnitudes.
Our final shortlist of BGM stars was obtained by keeping only BGM stars that were deemed close analogues of \textit{Kepler} stars. This selection is detailed in Section \ref{sec:kepler}.

\subsection{Exoplanet model} \label{sec:plan-model}

Having generated a catalogue of host stars, we now populate planets around them. Planetary physical characteristics are assigned using parameterised distribution functions. We determine the number of planets around each star using a weighted planet occurrence rate, which depends on the stellar mass and type, as described later in this section. Subsequently, these planets are assigned physical parameters required to compute their detectability for a given method. An overview of the TAED forward model and retrieval workflow is shown in Figure~\ref{fig:taed_flow}. In this paper, we are testing the approach using only simulated transit samples, and so we provide details on the generation of observables relevant to transit detection.

Motivated by \cite{Pascucci2018AStars}, we adopt the planet--host mass ratio as the primary and universal descriptor of planet bulk characteristics that, together with host mass, controls planet occurrence.

Accordingly, we assume a two-stage broken power law for the number $N$ of planets with planet-star mass ratio, $q$, of the form
\begin{equation}
    \frac{\mathrm{d} N}{\mathrm{d}\log q} \propto \left( \frac{q}{q_{\rm br}} \right)^{n} ~\text{where~} n = \begin{cases}
    n_1 & \quad (q_1 \leq q \leq q_{\rm br}) \\
    n_2 & \quad (q_{\rm br} < q \leq q_2)
    \end{cases},
\label{eq:q_dist}
\end{equation}
where \textit{$n$} is a power-law exponent. 

Defining $\beta$ as the ratio of the sum of masses of all planets around a host star to the mass of the star, we have
\begin{equation}
\beta = \langle N_p \rangle\frac{\langle M_{\mathrm{p}} \rangle}{M_{\star}} = \langle N_p \rangle \langle q \rangle,
\label{eq:beta}
\end{equation}
where $\langle \ldots \rangle$ denotes averaging within a given host system and $N_p$ is the number of planets per star. From Equation \eqref{eq:q_dist}, the average mass ratio is
\begin{equation}
    \langle q \rangle = \frac{\frac{\ts 1}{\ts n_1+1}q_{\rm br}^{1-n_1}(q_{\rm br}^{n_1+1}-q_1^{n_1+1})+\frac{\ts 1}{\ts n_2+1}q_{\rm br}^{1-n_2}(q_2^{n_2+1}-q_{\rm br}^{n_2+1})}{\frac{\ts 1}{\ts n_1}q_{\rm br}^{1-n_1}(q_{\rm br}^{n_1}-q_1^{n_1})+\frac{\ts 1}{\ts n_2}q_{\rm br}^{1-n_2}(q_2^{n_2}-q_{\rm br}^{n_2})}.
    \label{eq:qav}
\end{equation}

By fixing a universal value for $\beta$, we compute $\langle N_p \rangle = \beta / \langle q \rangle$ and then, for each star, select the number of planets per star, $N_p$, to be a Poisson deviate about $\langle N_p \rangle$. Whilst the model may predict multiple planets exist around each host, we do not model multi-planet transit detections. The role $N_p$ plays is in determining the overall yield of single-planet transits. The larger the planet multiplicity, the larger the overall expected yield will be of single-planet transit lightcurves. 

Like $q$, the planet--host separation, $a$, is allowed to follow a two-stage power-law distribution of the form
\begin{equation}
    \frac{\mathrm{d} N}{\mathrm{d}\log a} \propto \left( \frac{a}{a_{\rm br}} \right)^{y} ~\text{where~} y = \begin{cases}
    y_1 & \quad (a_1 \leq a \leq a_{\rm br}) \\
    y_2 & \quad (a_{\rm br} < a \leq a_2).
    \end{cases}
    \label{eq:sep}
\end{equation}
For adopted values of $\beta$, $n$, $q_{\rm max}$, $q_{\rm min}$, and $q_{\rm br}$ we can use Equations \eqref{eq:qav} and \eqref{eq:beta} to determine $N_p$ for a given system and then use Equations \eqref{eq:q_dist} and \eqref{eq:sep} to randomly generate $N_p$ pairs of $q$ and $a$. $q$ is converted to planet mass via $M_p = q M_{\star}$ and then to radius via the mass-radius-temperature relations in \cite{2023arXiv231016733E}. Planet temperature is computed using the BGM output values for stellar effective temperature, together with $a$, to determine equilibrium temperature for a fixed Bond Albedo of 0.3. Finally, the period of the exoplanet, $P$, is calculated directly from $a$ and $M_{\star}$ via Kepler's 3rd law.

We make a number of strong assumptions for the initial tests presented in this paper. Firstly, we assume that planets lie on circular orbits. Whilst eccentricity, $e$, is typically poorly constrained from a transit light curve in isolation (since the observed duration depends on $e$ and longitude of periastron for the planet's
orbit, $\omega$, \citep{2007PASP..119..986B} and is degenerate with the stellar density \citep{Sandford_2017}), it can still modify the transit duration and the geometric transit probability. The main effect of ignoring eccentric orbits on demographic modelling of transits would be a potentially biased recovery of $\beta$, which governs planet multiplicity. We also do not model the effect of mutual inclination and instead assume all planets to be coplanar within a given system. Mutual inclination increases the probability of observing at least one transit around a given system, so ignoring this can also bias recovery of $\beta$.

In principle, it is straightforward to add both eccentricity and mutual inclination distributions into the TAED framework. Ultimately, as the number of fit parameters increases, the recovery precision will be governed by the size and quality of the dataset, and the extent to which the data provides complementary sensitivity to parameters.

This is where the addition of a microlensing exoplanet sample may be helpful. In the case of planets detected through microlensing, the distribution of eccentricities will imprint directly upon the distribution of projected planet--host separation. A joint demographic analysis of transit and microlensing samples may well therefore be able to recover separately the planet multiplicity and eccentricity distribution. The inclusion of mutual inclination is likely to be important for observed transit yield but unimportant for microlensing. This complementary sensitivity will therefore play a crucial role within a TAED-based analysis.

As we are focusing in this paper only on transiting planets, we will leave to future papers investigations of how joint transit and microlensing samples may be able to constrain eccentricity and mutual inclination distributions.

\section{Simulating a \textit{Kepler}-like transit sample}
\label{sec:kepler}

The \textit{Roman} transit sample is expected to be 1.5-2 orders of magnitude larger than that obtained by \textit{Kepler}. It will comprise transits observed over Galactic distances, not just within 1~kpc of the Sun.
Nonetheless, the \textit{Kepler} exoplanet sample is the largest to date obtained by a single observatory, from the ground or space, via any detection method. Additionally, it has a very well characterised detection efficiency \citep{2017ksci.rept...19B}. It therefore provides the best available analogue for testing a TAED framework applicable to \textit{Roman}.

To simulate a sample indicative of one observed by a mission like \textit{Kepler}, we need to assert an equivalence between a BGM simulated host star and a real star within the \textit{Kepler} input catalogue. This matching was achieved by assigning a similarity score
\begin{equation}
S=\sqrt{\left(\frac{\Delta T_{\star}}{T_{\star}}\right)^2 +\left(\frac{\Delta R_{\star}}{R_{\star}}\right)^2 +\left(\frac{\Delta K_p}{K_p}\right)^2 +\left(\frac{\Delta M_{\star}}{M_{\star}}\right)^2  },
\label{eq:similar_star}
\end{equation}
where small $S$ is better and $S = 0$ would indicate a perfect match.
The numerators in Equation \eqref{eq:similar_star} refer to the difference in the value of the parameter of the star from the BGM sample to that from the \textit{Kepler} input catalogue (i.e $\Delta T_{\star}= T_{\star, \mathrm{BGM}}- T_{\star, \mathrm{Kepler}}$, and so on). Denominators refer to the parameters from the \textit{Kepler} catalogue. The parameters $R_\star$, $T_\star$, $K_P$, and $M_\star$ were selected as a compact proxy for the stellar dependence entering the KeplerPORTs contours \citep{2017ksci.rept...19B}. %they are the stellar parameters used by KeplerPORTs \citep{2017ksci.rept...19B} to determine detection efficiency.  

We ensure that matched stars are located within the same \textit{Kepler} detector module, and we discard BGM stars that have $S> 0.15$. Not all stars that \textit{Kepler} is able to detect are used within KeplerPORTs. So the number of matching BGM stars does not bear direct correspondence to the number within the \textit{Kepler} catalogue. Because of this, our simulated yields are rescaled to match what they would be if the simulated and real samples were of the same size.

At this point, we have a TAED transit simulation framework to compute observable transit yields for a given set of planet model parameters. But what we ultimately wish to do is to perform model retrieval -- i.e. be able to recover the planet model parameters that provide the best match to an observed dataset. This is what we now turn to.

\begin{figure*}
	\includegraphics[width=2\columnwidth]{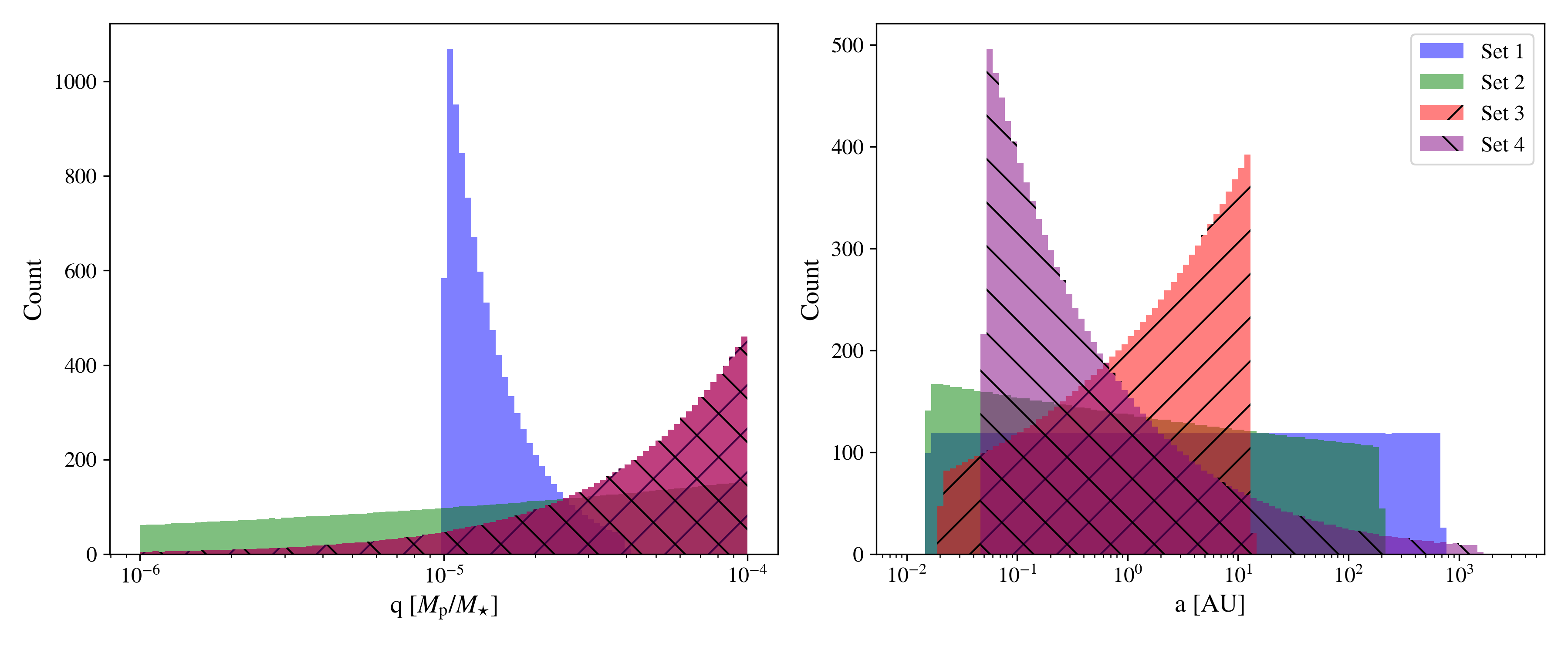}
    \caption{The distribution of planet-to-star mass ratio and semi-major axis in the different injected sets listed in Table~\ref{tab:self-test}. Note that, in the left panel, the $q$ distributions for Sets 3 and 4 are identical and so overlay on top of each other. These distributions define the synthetic planetary populations used to test the retrieval framework. The parameter ranges are chosen to span a wide variety of physical scenarios, including compact and extended planetary systems, as well as narrow- and broad-range of mass-ratios of planets. This ensures that the retrieval framework is tested across a broad and realistic spectrum of exoplanetary architectures.}
\label{fig:injected_distributions}
\end{figure*}

\section{Planetary Model retrieval}
\label{sec:retrieval}

To test the sensitivity and efficacy of a TAED retrieval framework, we perform injection tests where we use the planet-host model in Section~\ref{sec:taed} with a known set of model parameters to generate a synthetic observed dataset. We then perform a retrieval on the synthetic dataset to test the recovery of the parameters. 

For our initial tests, we use a simplified set of planetary models with single slope power laws in $q$ and $a$. So we set $n = n_1 = n_2$ in Equation \eqref{eq:q_dist} and $y = y_1 = y_2$ in Equation \eqref{eq:sep}. But even these simplified injection sets provide 7 parameters to retrieve: the minimum and maximum of planet-host mass ratio $q$ ($q_1$, $q_2$), along with the $q$ distribution slope $n$; the minimum and maximum for host separation $a$ ($a_1$, $a_2$), along with the $a$ distribution slope $y$; and $\beta$ which is the average mass per host in planets, normalised to the host mass. We defer to Section \ref{sec:kepler-params} to test recovery for the full 11-parameter models that employ broken power law distributions for $q$ and $a$.

We select 4 different injected models to represent different ground truths. Set 1 is chosen to span a relatively narrow range of $q$, but with a wide and flat distribution in $a$. Set 2 spans a broad range in $q$ with a modest slope favouring larger $q$. It has a more restricted span in $a$ than Set 1. Set 3 has the same span in $q$ as Set~2 but with a larger slope. It also has a very restricted range in $a$. Set 4 has the same $q$ model as Set 3 but has the broadest range of $a$ of all of the models. In each case, a value of $\beta$ is chosen so that the predicted average total number of planets per host is comparable to the number of Solar System planets. 

For each model set, we ran the simulation to provide an ``observed'' yield in the limits of the perfect detection efficiency. Given differences in the models, this means the underlying number of generated planets differs markedly between the sets. We list our injected model parameters in Table \ref{tab:self-test}, with the corresponding distributions of $q$ and $a$ shown in Figure \ref{fig:injected_distributions}. 

The simulations are used to generate a two-dimensional histogram of planet periods and radii for the injected parameter vector. This represents our synthetic ``observed'' dataset. For each trial model, we obtain an equivalent 2D histogram, representing the model prediction. The likelihood of a model is determined by the probability, $\Pr$, that the 2D histograms of the observation and model match, via: 
\begin{equation} 
\mathcal {L} = \prod _{i,j}^{\text{bins where }N_{ij}^{\text{obs}}>0}\Pr(N_{ij}^{\text{obs}}\mid N_{ij}^{\text{mod}}), 
\label{eq:like}
\end{equation}
where $N_{ij}^{\text{obs}}$ gives the number of planets in the $i,j$ period-radius bin for the synthetic observed dataset, and $N_{ij}^{\text{mod}}$ gives the same for the model prediction. 

When attempting to retrieve results from the actual \textit{Kepler} data, normally the model prediction would be multiplied through by the KeplerPORTs detection efficiencies to determine $N^{\text{mod}}$ in Equation \eqref{eq:like}, so that $N^{\text{mod}}$ could be compared directly to the observed sample represented by $N^{\text{obs}}$. In this case, for the correct model, we would expect $N^{\text{obs}}$ to be Poisson distributed about $N^{\text{mod}}$ within each period-radius bin. However, this approach requires efficiency correction to every likelihood evaluation, which is computationally very expensive. 

An alternative approach is to instead apply the KeplerPORTs efficiencies to the observed sample to determine $N_{ij}^{\text{obs},\epsilon} = \sum_k 1/\epsilon_{ijk}$, where the KeplerPORTs efficiency, $\epsilon_{ijk}$, is extracted for every observed transit belonging to period-radius bin $ij$. Here, index $k$ spans all transit properties over which the detection efficiency is tabulated, other than planet period and radius. $N_{ij}^{\text{obs},\epsilon}$ then provides an estimate of the yield expected for the survey were it to have had perfect detection efficiency, allowing it to be compared directly to the \textit{uncorrected} model prediction, $N_{ij}^{\text{mod}}$. In this case, we only need to apply efficiency correction once, to the observed sample, and not to each model likelihood evaluation. This is the approach we take, though it should be stressed that it has some limitations. 

Firstly, unlike $N^{\text{obs}}$, $N^{\text{obs},\epsilon}$ is non-integer and therefore cannot be Poisson distributed about $N^{\text{mod}}$. Instead, the effect of $\epsilon$ is to amplify Poisson fluctuations in $N^{\text{obs}}$ by a factor $1/\epsilon$. Since the Poisson distribution mean and variance both scale linearly with $N_{\text{mod}}$, a $1/\epsilon$ noise amplification in the mean and variance of $N^{\text{obs}}$ should not significantly bias the retrieval, such that the distribution of $N^{\text{obs},\epsilon}$ can be considered \textit{quasi}-Poissonian about an uncorrected $N^{\text{mod}}$. Nonetheless, in regions of low detection efficiency (high noise amplification) $N^{\text{obs},\epsilon}$ becomes numerically unstable. This means that we should restrict likelihood evaluations to regions of planet-host parameter space where $\epsilon$ is comfortably above zero. This is straightforward to handle in the retrieval algorithm, meaning our simplified and much faster retrieval approach should still yield a reliable recovery for a survey that has good efficiency over at least some region of the discovery space.

Taking $N_{ij}^{\text{obs},\epsilon}$ to be quasi-Poisson distributed about $N_{ij}^{\text{mod}}$ for every period-radius bin, $i,j$, our retrieval aims to find the model parameter vector that maximises
\begin{equation}
    \log {\cal L} = \sum_{i,j}^{N_{ij}^{\text{obs},\epsilon} > 0} N_{ij}^{\text{obs},\epsilon} \log (N_{ij}^{\text{mod}}/N_{ij}^{\text{obs},\epsilon}) + N_{ij}^{\text{obs},\epsilon}- N_{ij}^{\text{mod}},
    \label{eq:loglike}
\end{equation}
where Stirling's approximation is used for $\log (N^{\text{obs},\epsilon}!)$. Note that, whist $N^{\text{obs},\epsilon}$! is undefined for non-integer $N^{\text{obs},\epsilon}$, Stirling's approximation is defined for non-integers. 

\section{Testing retrieval algorithms} \label{sec:retrieve-algo}

With an eye on the very large planet samples that are expected to be delivered by \textit{Roman}, we test four different retrieval algorithms for their efficiency and accuracy.
These include: nested sampling; random uniform sampling; a 2-stage machine learning implementation; and differential evolution \citep{storn_differential_1997}. 

\subsection{Nested sampling}
\label{sec:nested-sampling}
\begin{figure*}
	\includegraphics[width=2\columnwidth]{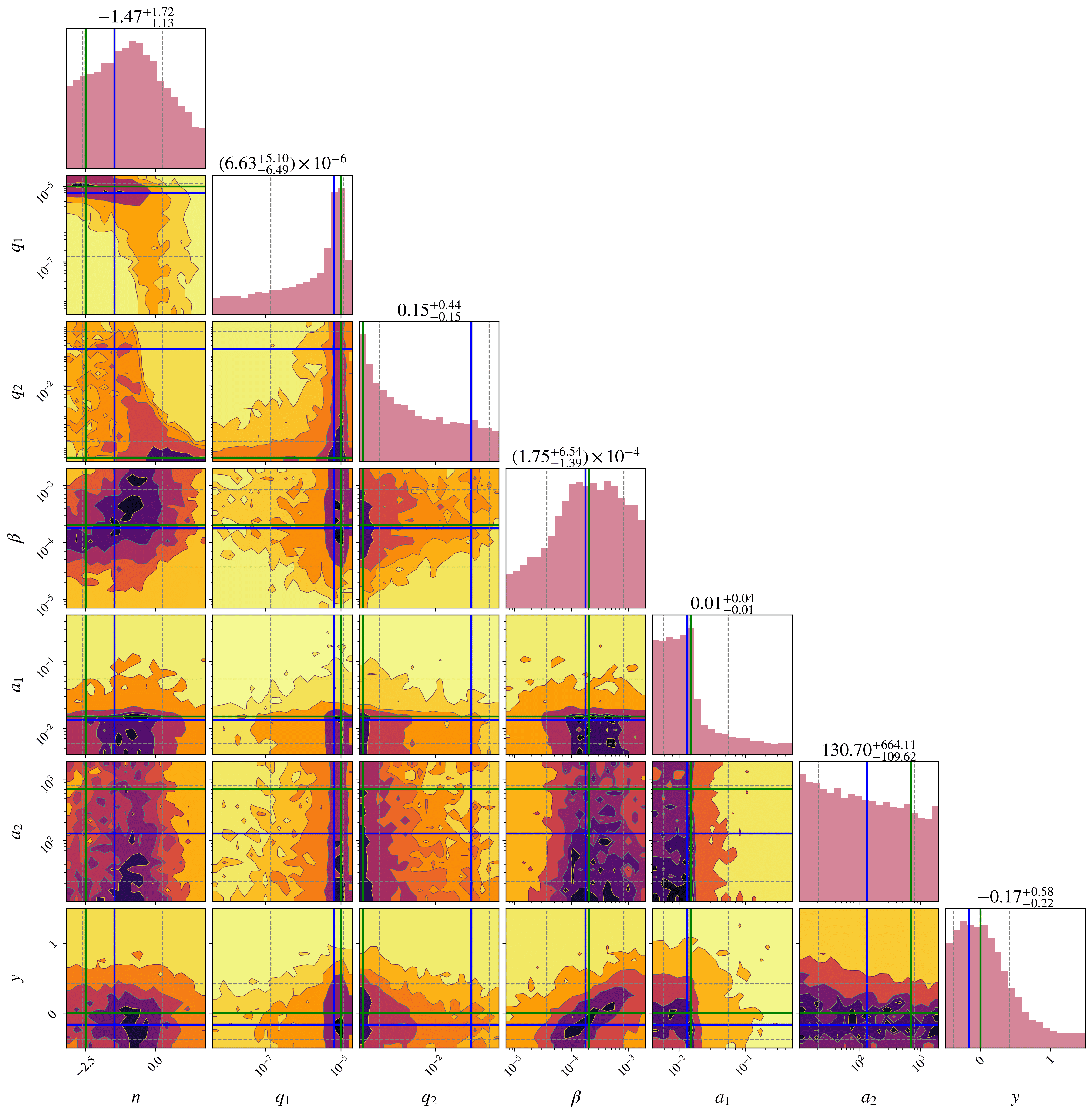}
    \caption{Cornerplot showing 1-D marginals (diagonals) and 2-D joint posteriors (off-diagonals) of samples of parameters while recovering parameters for Set 1 (broad $a$, peaked $q$) using \dyne{} nested sampling (NS). The multi-dimensional best-fit value is overplotted in blue colour, while the injected parameters are overplotted in green colour. The dashed lines show the upper and lower errors on the best-fit value.}
\label{fig:cornerplot_dynesty}
\end{figure*}

For nested sampling, we employed the widely used \dyne{} algorithm \citep{2020MNRAS.493.3132S}, though we also did some testing with \textsc{Ultranest} \citep{2021JOSS....6.3001B}, which was found to be consistent with but slower than \dyne{}. The main issue we encountered with \dyne{} was that after $\sim$1,000,000 calls, the remaining evidence typically converged to a value much larger than the default convergence threshold. We attribute this to the fact that both the model and observed samples are derived from discrete simulation data, meaning that both are intrinsically noisy, which hampers the retrieval of the optimal model. This behaviour was confirmed through testing the retrieval of simpler toy model distributions. The results from nested sampling are presented in Table \ref{tab:self-test}, and the posterior distribution for Set 1 (broad-$a$, peaked $q$)  is shown in Figure \ref{fig:cornerplot_dynesty}.  The results demonstrate the method's ability to recover input demographics, though with varying precision across parameters. In all the instances, the injected values lie within or close to the 1-sigma error interval. Some parameters (e.g., $n$ and $a_2$) show broader distributions, suggesting degeneracies or lower sensitivity in the dataset to those parameters. In Set 1, the possible $q$ range is quite narrow, spanning from $1\times10^{-5}$ to $4\times10^{-5}$. This limited dynamic range can reduce the sensitivity of the slope parameter $n$, as the model lacks sufficient leverage to distinguish between subtle variations. Moreover, the transit probability scales inversely with $a$, so as $a_2$ increases, there are more planets with a lower transit likelihood. This leads to a loss of sensitivity in the model's ability to constrain parameters associated with distant orbits. This is evident in the broader posterior distribution and reduced density in the scatter plots involving $a_2$. The limits $q_1$ and $q_2$ exhibit some negative correlation with $n$, with Pearson's coefficients of -$0.47$ and $-0.39$ respectively. When $n$ is increased, for a fixed $\beta$, it increases the proportion of high-$q$ planets. 

\begin{table*}
    \centering
    \caption[]{Injected (IN) and recovered parameters using four different retrieval methods applied to a 7-parameter planetary demographic model. The methods used are: \dyne{} nested sampling (NS); uniform random sampling (UR), accelerated uniform random sampling through a two-stage machine learning model (ML); and differential evolution based retrieval (DE). The last row shows the limits of corresponding priors. 
    The range of $\beta$ was chosen to yield between 1 and 15 planets per star on average, across the full range of other parameters. Parameters $n$ and $y$ were sampled from linear priors, whilst other parameters were sampled from log-uniform space. The timings are reported while using 60 CPU cores on an AMD Ryzen Threadripper PRO 3995WX, except for NS, where we found that using a single core was faster than multiple cores for our retrieval. For ML retrievals, we mention the timings to train the model (using 60 cores) and for likelihood prediction (using a single core).}
    \label{tab:self-test}
    \centering
    \rotatebox{90}{\begin{tabular}{llllllllll}
    \hline
        Set & ~ & $n$ & $q_1$ & $q_2$ & $\beta$ & $a_1$ & $a_2$ & $y$ & Runtime (s)\\ [.35em]\hline
        1 & IN & $-2.50$ & $1.00 \times 10^{-5}$ & $4.00 \times 10^{-5}$ & $2.00 \times 10^{-4}$ & $0.01$ & $700.00$ & $0.00$   \\ 
        ~ & UR & $-2.15^{+2.02}_{-0.69}$ & $(9.00^{+3.76}_{-8.45}) \times 10^{-6}$ & $(2.73^{+379.00}_{-2.26}) \times 10^{-4}$ & $(1.64^{+6.30}_{-1.00}) \times 10^{-4}$ & $(1.54^{+0.18}_{-0.91}) \times 10^{-2}$ & $97.60^{+627.12}_{-78.10}$ & $-0.04^{+0.32}_{-0.30}$  & 35,279  \\[.35em] 
        & ML & $-0.21^{+0.80}_{-1.68}$ & $(9.37^{+821.31}_{-8.11}) \times 10^{-8}$ & $(8.99^{+703.41}_{-8.43}) \times 10^{-4}$ & $(1.93^{+0.05}_{-1.77}) \times 10^{-3}$ & $(8.89^{+5.33}_{-3.73}) \times 10^{-3}$ & $11.63^{+240.38}_{-1.15}$ & $0.53^{+0.16}_{-0.74}$ & 19+37\\[.35em]
        & NS &  $-1.47^{+1.72}_{-1.13}$ & $(6.63^{+5.10}_{-6.49}) \times 10^{-6}$ & $0.15^{+0.44}_{-0.15}$ & $(1.75^{+6.54}_{-1.39}) \times 10^{-4}$ & $0.01^{+0.04}_{-0.01}$ & $130.70^{+664.11}_{-109.62}$ & $-0.17^{+0.58}_{-0.22}$& 243,696\\[.35em]
        & DE& $-2.67^{+1.16}_{-0.25}$ & $(1.01^{+0.10}_{-0.54}) \times 10^{-5}$ & $(3.18^{+59.39}_{-2.90}) \times 10^{-3}$ & $(1.43^{+1.87}_{-0.93}) \times 10^{-4}$ & $(1.49^{+0.58}_{-0.39}) \times 10^{-2}$ & $11.91^{+245.61}_{-1.32}$ & $0.02^{+0.24}_{-0.15}$ & 4,079
\\[.35em]\hline
        2 & IN & $0.20$ & $1.00 \times 10^{-6}$ & $1.00 \times 10^{-4}$ & $2.50 \times 10^{-4}$ & $0.01$ & $200.00$ & $-0.05$  \\ 
        ~ & UR & $0.06^{+0.62}_{-0.57}$ & $(7.82^{+19.49}_{-7.58}) \times 10^{-7}$ & $(1.53^{+13.71}_{-0.86}) \times 10^{-4}$ & $(2.53^{+7.89}_{-1.61}) \times 10^{-4}$ & $(1.52^{+0.18}_{-0.89}) \times 10^{-2}$ & $461.25^{+774.05}_{-431.97}$ & $-0.09^{+0.31}_{-0.27}$ & 35,426  \\[.35em]
        & ML & $-0.01^{+0.62}_{-0.48}$ & $(2.84^{+16.64}_{-2.71}) \times 10^{-7}$ & $(2.19^{+17.71}_{-1.49}) \times 10^{-4}$ & $(1.46^{+0.37}_{-1.23}) \times 10^{-3}$ & $(1.21^{+0.34}_{-0.63}) \times 10^{-2}$ & $10.06^{+215.10}_{-0.05}$ & $0.40^{+0.13}_{-0.60}$ & 19+37 \\[.35em] 
        & NS &  $0.12^{+0.48}_{-1.11}$ & $(7.31^{+28.03}_{-7.04}) \times 10^{-7}$ & $(1.10^{+19.72}_{-0.47}) \times 10^{-4}$ & $(2.78^{+8.08}_{-2.26}) \times 10^{-4}$ & $0.02^{+0.06}_{-0.01}$ & $25.50^{+406.69}_{-12.28}$ & $-0.05^{+0.38}_{-0.28}$& 360,040\\[.35em]
    &DE& $0.12^{+0.21}_{-0.62}$ & $(1.07^{+1.11}_{-0.72}) \times 10^{-6}$ & $(1.30^{+4.15}_{-0.36}) \times 10^{-4}$ & $(3.33^{+4.00}_{-1.73}) \times 10^{-4}$ & $(1.49^{+0.83}_{-0.40}) \times 10^{-2}$ & $180.63^{+483.95}_{-144.17}$ & $-0.02^{+0.23}_{-0.14}$ & 3,454

  \\[.35em] \hline
        3 & IN & $1.00$ & $1.00 \times 10^{-6}$ & $1.00 \times 10^{-4}$ & $5.00 \times 10^{-4}$ & $0.02$ & $13.00$ & $0.25$  \\
        ~ & UR & $0.96^{+0.51}_{-1.40}$ & $(6.13^{+4467.55}_{-1.55}) \times 10^{-9}$ & $(9.90^{+83.35}_{-3.69}) \times 10^{-5}$ & $(5.01^{+7.96}_{-3.64}) \times 10^{-4}$ & $(2.01^{+0.36}_{-1.33}) \times 10^{-2}$ & $26.71^{+405.49}_{-13.28}$ & $0.19^{+0.23}_{-0.46}$  & 35,378   \\[.35em] 
        ~& ML & $0.54^{+0.72}_{-1.12}$ & $(2.84^{+576.87}_{-2.10}) \times 10^{-8}$ & $(1.68^{+14.99}_{-0.92}) \times 10^{-4}$ & $(1.92^{+0.05}_{-1.68}) \times 10^{-3}$ & $(7.00^{+9.91}_{-2.22}) \times 10^{-3}$ & $11.27^{+268.80}_{-0.89}$ & $0.57^{+0.15}_{-0.78}$& 19+37 \\[.35em]
        & NS &  $0.90^{+0.39}_{-1.35}$ & $(1.16^{+6.60}_{-1.13}) \times 10^{-6}$ & $(1.21^{+10.42}_{-0.37}) \times 10^{-4}$ & $(1.27^{+0.46}_{-1.05}) \times 10^{-3}$ & $(5.66^{+13.35}_{-1.19}) \times 10^{-3}$ & $12.34^{+245.85}_{-1.66}$ & $0.52^{+0.47}_{-0.57}$& 303,082\\[.35em]
        &DE&$0.96^{+0.21}_{-0.71}$ & $(5.30^{+22.91}_{-5.04}) \times 10^{-7}$ & $(1.11^{+2.18}_{-0.25}) \times 10^{-4}$ & $(7.32^{+4.83}_{-5.05}) \times 10^{-4}$ & $0.02^{+0.02}_{-0.01}$ & $22.48^{+142.05}_{-8.61}$ & $0.27^{+0.18}_{-0.22}$ & 3,294
\\[.35em]\hline
        4 & IN & $1.00$ & $1.00 \times 10^{-6}$ & $1.00 \times 10^{-4}$ & $5.00 \times 10^{-4}$ & $0.05$ & $1500.00$ & $-0.40$  \\ 
        ~ & UR & $1.18^{+0.38}_{-1.60}$ & $(5.28^{+694.43}_{-4.36}) \times 10^{-8}$ & $(1.03^{+5.72}_{-0.38}) \times 10^{-4}$ & $(5.53^{+8.14}_{-3.13}) \times 10^{-4}$ & $0.05^{+0.01}_{-0.04}$ & $45.63^{+489.97}_{-30.07}$ & $-0.44^{+0.46}_{-0.04}$ & 35,666  \\ [.35em]
        & ML & $0.03^{+0.98}_{-1.13}$ & $(3.06^{+557.59}_{-2.31}) \times 10^{-8}$ & $(2.93^{+40.51}_{-2.06}) \times 10^{-4}$ & $(1.89^{+0.07}_{-1.43}) \times 10^{-3}$ & $0.03^{+0.02}_{-0.02}$ & $10.10^{+263.09}_{-0.07}$ & $0.54^{+0.14}_{-0.80}$ & 19+37 \\[.35em]
        & NS &  $1.07^{+0.25}_{-0.79}$ & $(5.32^{+158.02}_{-4.43}) \times 10^{-8}$ & $(9.67^{+10.12}_{-1.86}) \times 10^{-5}$ & $(5.13^{+5.54}_{-3.60}) \times 10^{-4}$ & $(4.99^{+0.31}_{-3.37}) \times 10^{-2}$ & $1306.09^{+430.15}_{-1264.71}$ & $-0.36^{+0.37}_{-0.09}$ & 446,420\\[.35em]
        &DE&$0.97^{+0.22}_{-0.71}$ & $(8.86^{+25.98}_{-8.63}) \times 10^{-7}$ & $(1.03^{+1.68}_{-0.20}) \times 10^{-4}$ & $(5.37^{+2.69}_{-3.11}) \times 10^{-4}$ & $0.05^{+0.01}_{-0.02}$ & $68.47^{+428.94}_{-44.94}$ & $-0.35^{+0.33}_{-0.08}$ & 2,883 
\\[.35em]\hline
        Priors &  & $[-3.2, 1.8]$ & $[4\times 10^{-9}, 2\times 10^{-5}]$ & $[3\times 10^{-5}, 1.23]$ & $[7\times 10^{-6}, 2\times 10^{-3}]$ & $[0.004, 0.5]$ & $[10, 2000]$ & $[-0.5, 1.5]$ \\
        \hline
    \end{tabular}}
\end{table*}

In addition to these sensitivity-driven degeneracies, the quoted intervals in Table \ref{tab:self-test} are further broadened by the compressed likelihood statistic, stochastic forward modelling, and prior/parameterisation effects. The likelihood is evaluated on a discretised $(P,R_p)$ histogram [Equation \eqref{eq:loglike}], which compresses the planet catalogue to counts per period--radius bin and discards intra-bin behaviour. This compression means that distinct demographic parameter vectors can map to very similar binned $(P,R_p)$ densities, yielding extended degeneracy directions and non-Gaussian marginals. Moreover, the forward model is stochastic at multiple stages, including drawing multiplicities via a Poisson deviate about $\langle N_p\rangle$, intrinsic scatter in the adopted mass--radius relation, and geometric transit observability; the likelihood surface can exhibit small-scale jitter that can broaden intervals.
Finally, several parameters are sampled with log-uniform priors spanning many decades (e.g.\ $q_1$, $q_2$, $\beta$, $a_1$, $a_2$), so reporting uncertainties in linear space can therefore produce highly asymmetric intervals and large relative errors even when the constraint in $\log$ space is comparatively modest.

\subsection{Random uniform sampling}
\label{sec:random-uniform}
Uniform random sampling represents a rather crude and inefficient retrieval method, compared to the other three approaches we test. However, it forms the basis for one of the other methods we employ, namely the two-stage machine learning approach. For this reason, its inclusion provides a good benchmark for the speed and accuracy of the other methods.

\begin{figure*}
	\includegraphics[width=2\columnwidth]{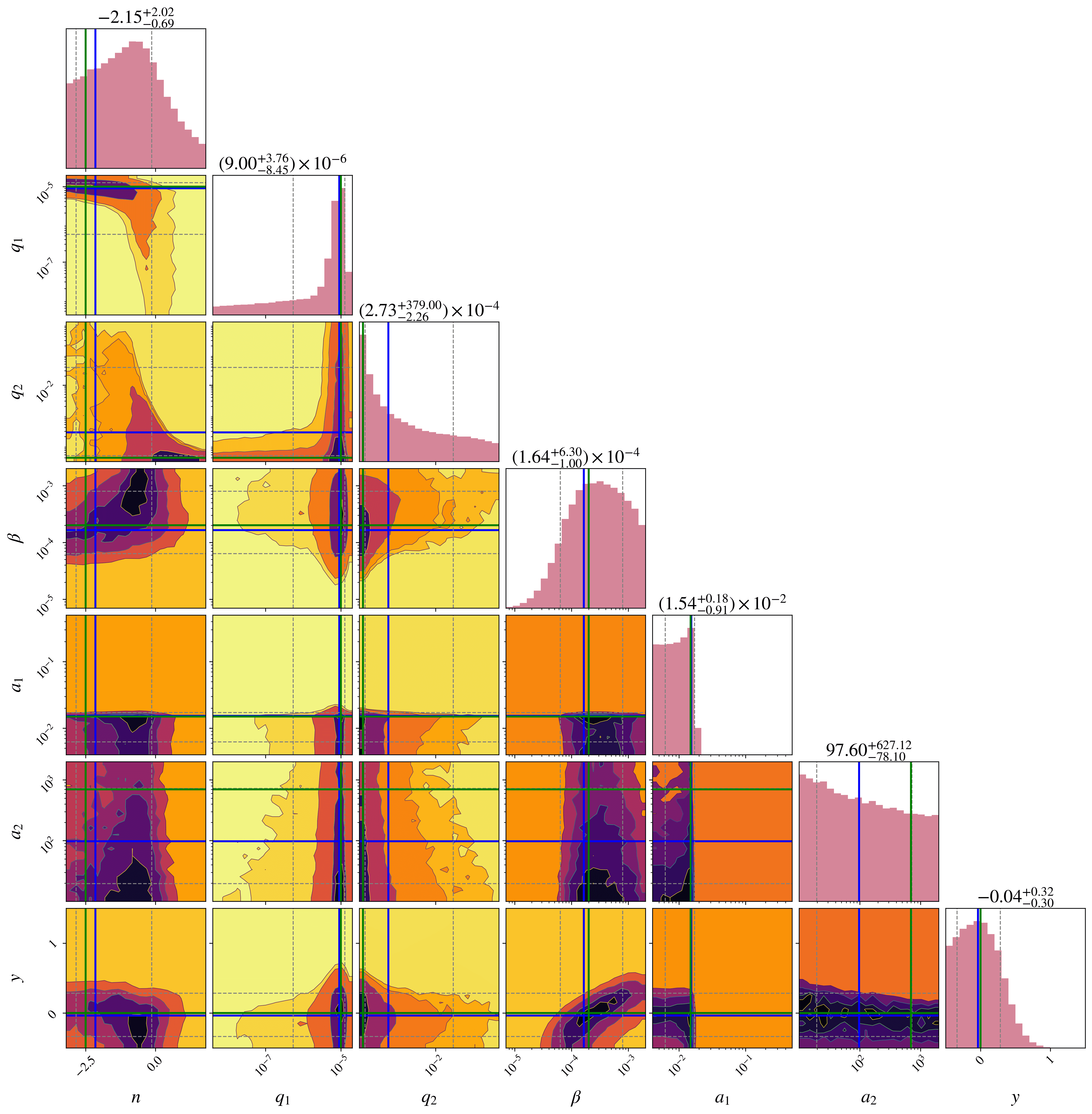}
    \caption{Cornerplot showing 1-D marginals (diagonals) and 2-D joint posteriors (off-diagonals) of samples of parameters while recovering parameters for Set 1 using the random uniform sampling method (UR). The multi-dimensional best-fit value is overplotted in blue colour, while the injected parameters are overplotted in green colour. The dashed grey lines show the upper-error and the lower-error on the best-fit value. The samples were uniformly distributed over the displayed parameter space, hence the density corresponds to the weights calculated in Section \ref{sec:random-uniform}.}
\label{fig:cornerplot}
\end{figure*}

We generated 10,000,000 parameter vectors sampled randomly from the uniform distribution of priors shown in Table \ref{tab:self-test}. Then we calculated the corresponding log-likelihood values. As the samples are distributed uniformly, the density of points in the posterior distribution is also uniform, so they must be weighted to obtain a meaningful posterior distribution. We calculate the weights, $w$, of each sample $i$ as
\begin{equation}
w_i \propto \exp\left(\alpha \left(\text{log}\mathcal {L}_i - \max(\text{log}\mathcal {L})\right)\right).
\label{eq:weights}
\end{equation}
We applied a scale factor $\alpha = 10^{-4}$ to suppress the dynamic range in ${\cal L}$ and mitigate against numerical underflow. We present the results from these tests in Table \ref{tab:self-test}, and show the posterior plots for Set 1 in Figure \ref{fig:cornerplot}. Similar to the retrieval with \dyne{} in Section \ref{sec:nested-sampling}, we see that the injected values are typically within a 1-sigma interval of the retrieved values, and the posterior distribution is analogous to that found from nested sampling. It presents a simple and robust solution for benchmarking, but it is not scalable.

\subsection{Two-stage machine learning}
\label{sec:ml}
\begin{figure}
	\includegraphics[width=\columnwidth]{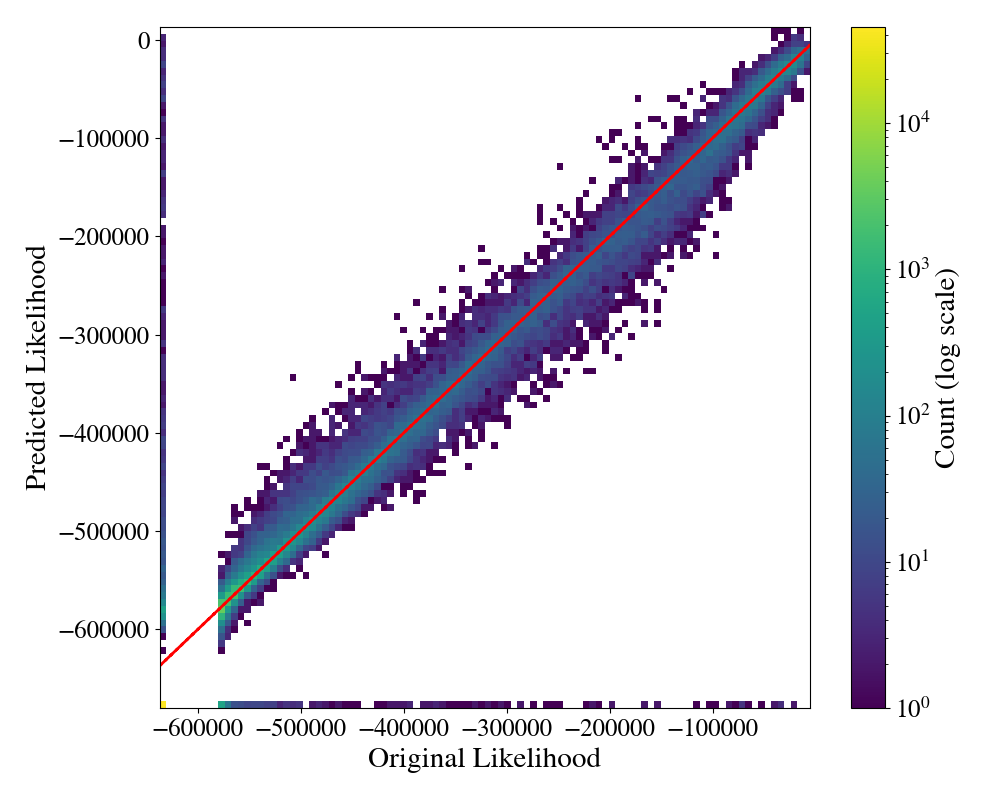}
    \caption{Figure showing the comparison of predicted likelihood using the 2-stage machine learning model for Set 1. The column of points on the left-most edge (false positives) of the original likelihood axis and a row of points on the bottom-most edge (false negatives) of the predicted likelihood axis correspond to 2.2\% of all data points. These points originate from misclassification in the first stage of the machine learning model.}
\label{fig:2stage-ml}
\end{figure}

\begin{figure*}
	\includegraphics[width=2\columnwidth]{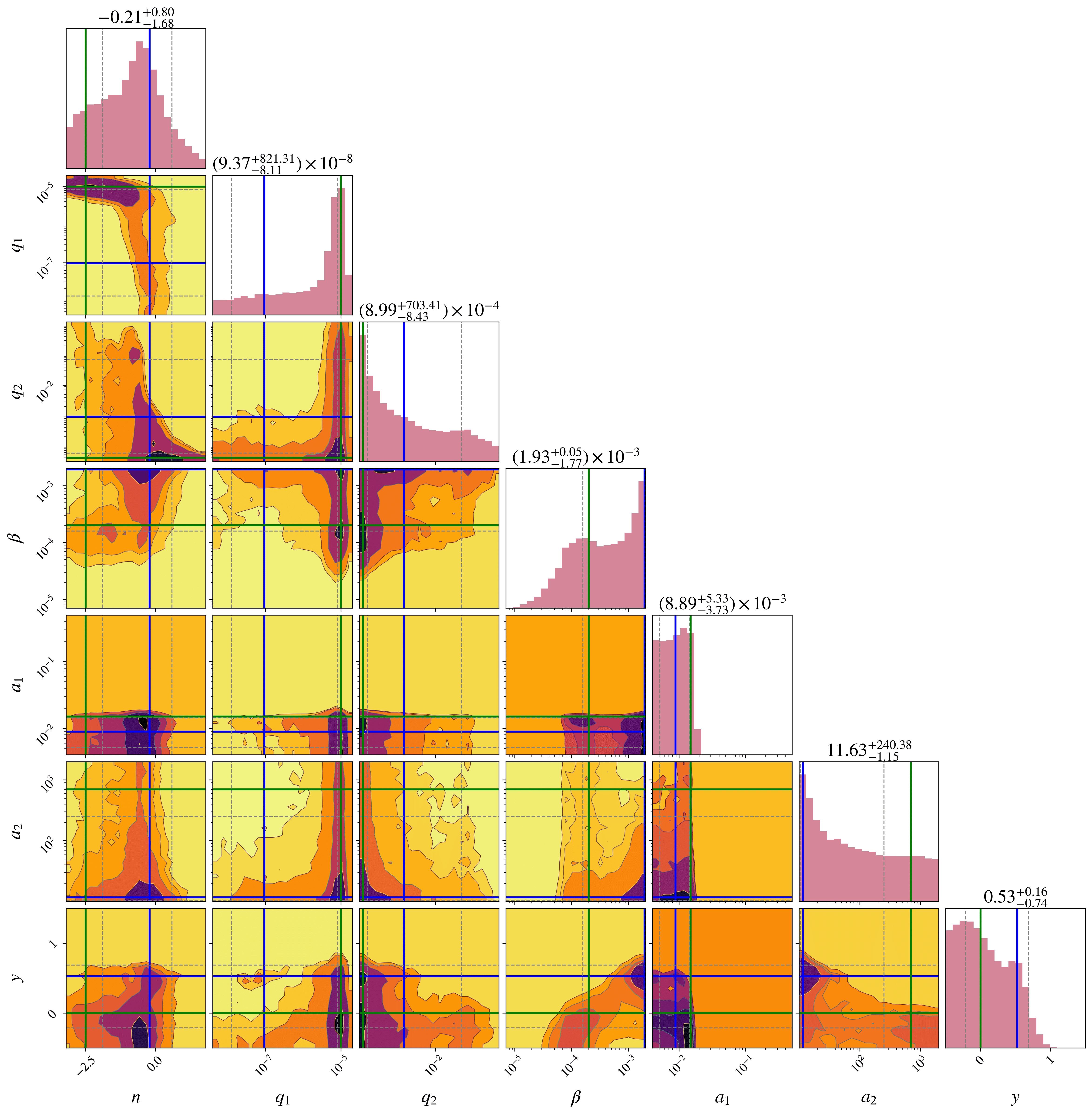}
    \caption{Cornerplot showing 1-D marginals (diagonals) and 2-D joint posteriors (off-diagonals) of samples of parameters while recovering parameters for Set 1, using the 2-stage machine learning model to predict log-likelihoods. The multi-dimensional best-fit value is overplotted in blue colour, while the injected parameters are overplotted in green colour. The dashed grey lines show the upper-error and the lower-error on the best-fit value. The samples were uniformly distributed over the displayed parameter space, hence the density corresponds to the weights calculated in Section \ref{sec:random-uniform}. }
\label{fig:cornerplot_ml}
\end{figure*}

In order to accelerate our likelihood evaluations, we prepared a 2-stage machine learning model to predict the likelihoods of parameter vectors. For a test dataset of exoplanets, we calculated the likelihoods corresponding to 250,000 random uniform samples for parameter vectors, with diminishing returns found for larger samples. Additional features were included in each sample through second-order combinations of all parameters in the samples. Around 46\% of the samples resulted in no exoplanets and so are penalised in the likelihood calculation. As these samples are not highly localised in parameter space, and given their potential to bias the prediction model, we create a Random Forest Classifier as a first stage to identify such samples. The second stage involves a Light Gradient Boost regression model \citep{zhang2017gpuaccelerationlargescaletreeboosting} to predict the likelihood of samples. For this stage, the likelihoods were standardised using a Z-score to avoid feature dominance simply through having a larger numerical range. We show the results from our prediction model in Figure \ref{fig:2stage-ml}. On the test dataset, the classifier model had an F1 score of 0.98, and the $R^2$ value of the regressor model was 0.99. The $R^2$ of the combined 2-stage solution was 0.94. Using this 2-stage machine learning model, we proceeded to generate log-likelihood predictions for 10,000,000 random samples. The weights of the sampled points are assigned using Equation \eqref{eq:weights}. The results from these analyses are shown in Table \ref{tab:self-test}, and the posterior plot for Set 1 is presented in Figure \ref{fig:cornerplot_ml}. The cornerplot shows that this method is capable of replicating key features of the posterior distribution seen from nested sampling retrieval, including the joint distributions between $n$ and $q_1$, and $n$ and $q_2$. However, when comparing with the retrievals from NS and UR, the marginalised distribution for $\beta$ shows a preference for higher values. This method is significantly faster than all other methods investigated, but presents a risk of producing unphysical high-likelihood islands. 

\subsection{Differential evolution to retrieve parameters}
\label{sec:diff-evo}

The final retrieval method we consider is differential evolution (DE) \citep{storn_differential_1997, 2020SciPy-NMeth}. DE is a stochastic method to find the global minimum of a function. It is designed to be efficient even when both the data and models are discrete, and is easy to highly parallelise. DE starts with an initial population of candidate parameter vectors and, for every candidate, creates a new trial candidate vector based on mixed and mutated combinations of existing vectors. A trial candidate vector succeeds its predecessor if it has a higher likelihood. The algorithm continues with a new generation of trial vectors and halts only when
\begin{equation}
    \sigma(E) \leq A_{\text{tol}} + R_{\text{tol}} \cdot |\mu(E)|,
    \label{eq:DE-halt}
\end{equation}
\noindent where $E = -\mathcal{L}$ are the ``energies'' of the samples from the current generation of candidate parameter vectors, and $\sigma(E)$ and $\mu(E)$ are the standard deviation and mean of $E$. $R_{\text{tol}}$ and $A_{\text{tol}}$ are relative and absolute tolerances, both of which are set to their default values of 0.01 and 0, respectively. The idea is that through mixing and mutation of parameter vectors, each generation of trial candidate samples will have a tendency to migrate towards an increasingly localised parameter region centred around the model of highest likelihood. Equation \eqref{eq:DE-halt} uses a measure of the tightness of this localisation as the basis for the halting condition. 

Given the smooth forms of our planet's demographic distributions, we choose the Sobol sampling method \citep{Sob67} to seed the initial generation of parameters. Sobol sampling is efficient at maximising the coverage of the parameter space, whilst avoiding clustered sampling encountered in random sampling, which would only decrease the convergence efficiency of the DE method. We set the initial population to be 8192. 

To generate a candidate sample in each iteration, we used the \texttt{randtobest1bin} method \citep{Qiang2014AUD}, which has been shown to perform better than other methods across diverse problem types and distributions \citep{Jeyakumar_2011}. For each candidate parameter array, $\mathbf{b}$, in the parent population, this strategy generates an intermediate candidate parameter array as 
\begin{equation}
    \mathbf{b}^{\prime}=\mathbf{x}_{r_0}+F \cdot\left(\mathbf{x}_0-\mathbf{x}_{r_0}+\mathbf{x}_{r_1}-\mathbf{x}_{r_2}\right),
\end{equation}
where $\mathbf{x}_{r_0}$, $\mathbf{x}_{r_1}$, and $\mathbf{x}_{r_2}$ are 3 random candidate parameter arrays drawn from the parent population, and $\mathbf{x}_0$ is the parameter array with the best likelihood from the current generation of candidate vectors. $F$ is the mutation parameter, which is sampled randomly for each generation from the interval $[0.5, 1)$. The parameter array for a new trial candidate, $\mathbf{b}_{t}$, is assigned by randomly selecting its elements from among those used for $\mathbf{b}^{\prime}$ or $\mathbf{b}$. It is ensured that at least one parameter is selected from $\mathbf{b}^{\prime}$. This is implemented by randomly choosing a parameter on each instance and then using the corresponding value from $\mathbf{b}^{\prime}$ for the trial candidate.  $\mathbf{b}$ is discarded in favour of $\mathbf{b}_{t}$ if $\mathbf{b}_{t}$ has a higher likelihood. In this way, a new successor generation of candidate parameter vectors is created.

\begin{figure*}
	\includegraphics[width=2\columnwidth]{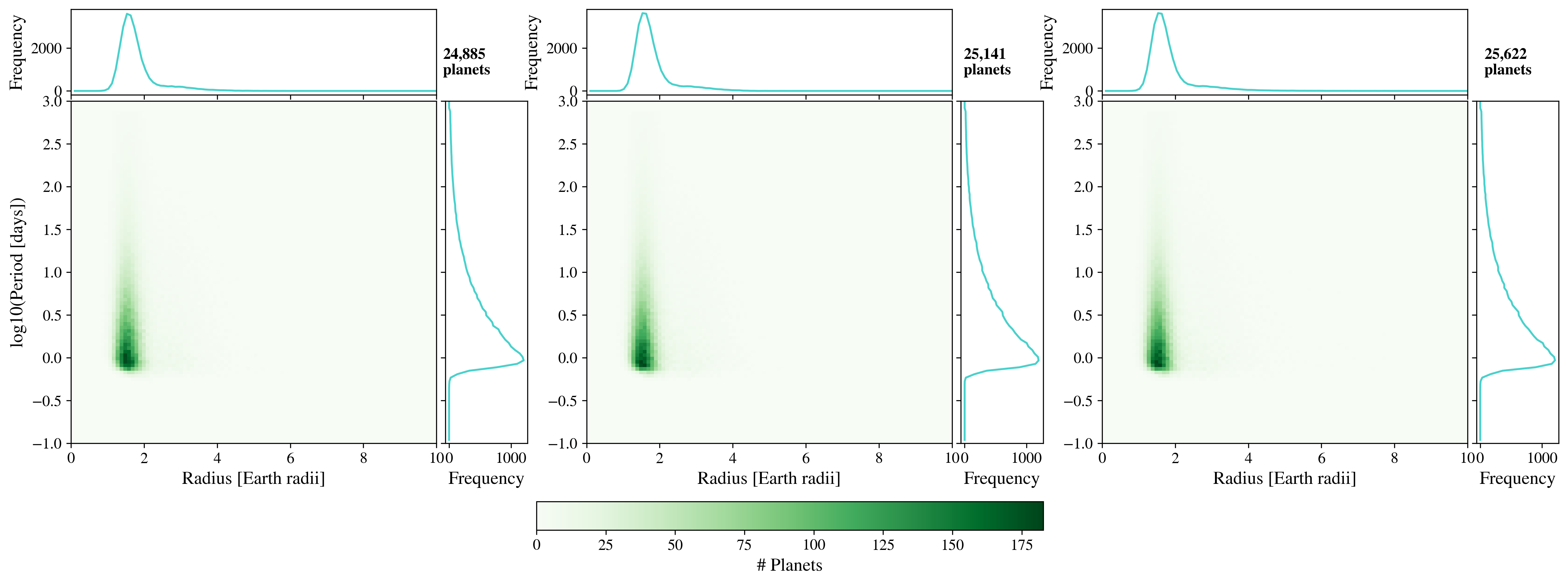}
    \caption{The figure shows the distribution of the injected dataset of exoplanets for Set 1, on the left and the corresponding recovered planets using differential evolution on the middle and right. In the middle panel, we only plot the bins where the injected distribution had non-zero planets. In the rightmost panel, we plot all the bins from the recovered population. The colourmap shows the number of planets on the period-radius axes. The axes are accompanied by 1-D histograms of radius
and period separately. In the top right corner of each 2-D histogram, the total number of the total planets is written. }
\label{fig:DE1}
\end{figure*}

\begin{figure*}
	\includegraphics[width=2\columnwidth]{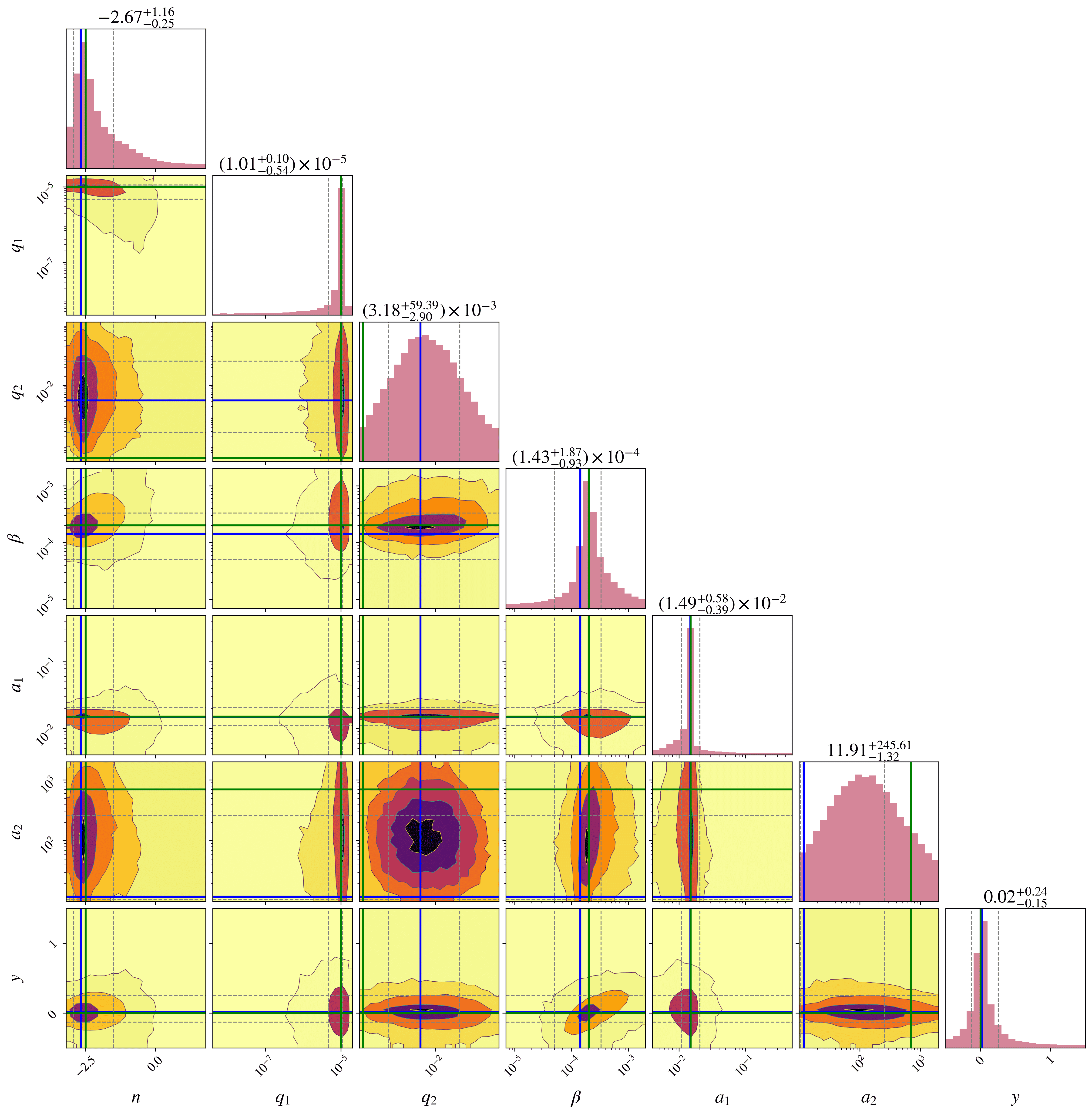}
    \caption{Cornerplot showing 1-D marginals (diagonals) and 2-D joint posteriors (off-diagonals) of samples of parameters while recovering parameters for Set 1 using differential evolution. The multi-dimensional best-fit value is overplotted in blue colour, while the injected parameters are overplotted in green colour. The dashed grey lines show the upper-error and the lower-error on the best-fit value.}
\label{fig:cornerplot_de}
\end{figure*}

The results of this method are summarised in Table \ref{tab:self-test}, and the comparison of population for Set 1 is shown in Figure \ref{fig:DE1}. The cornerplot for Set 1 is shown in Figure \ref{fig:cornerplot_de}. The posterior distributions from DE are more clearly delineated than in other retrievals, but they generally show good recovery of the injected parameters, which is consistent with the width of the posteriors. DE presents the best tradeoff between speed and accuracy of retrievals, which is efficiently parallelisable and scalable. However, DE is a stochastic global optimisation method and, unlike nested sampling, it does not generate correctly weighted samples from the Bayesian posterior.
Uncertainties estimated from the spread of the converged DE population are therefore heuristic and can underestimate true credible intervals.

\section{Parameter retrieval using the real Kepler dataset}
\label{sec:kepler-params}

Having verified the performance of the recovery using simulated data, we now recover parameters from the exoplanet sample observed by \textit{Kepler} itself. 

As mentioned in Section~(\ref{sec:kepler}), comparison of simulated and real \textit{Kepler} star samples required using a rescaling factor to account for the discrepancy between the number of stars in the BGM and the number chosen for the \textit{Kepler} Input Catalogue. \textit{Kepler} observed 150,000 stars, whereas 59,734 stars from the BGM sample qualified as representative of the \textit{Kepler} input catalogue using the similarity factor $S$ in Equation~\eqref{eq:similar_star} with the criterion $S \leq 0.15$.  Our simulated yields are accordingly scaled up by a factor 2.51 to represent the \textit{Kepler} expected yield for a given model. The scaling factor uses the commonly quoted \emph{Kepler} target list as a convenient normalisation, but the ``effective'' sample used for completeness/occurrence work is typically smaller after quality cuts \citep{10.1093/mnras/sty3463}. The DR25 pipeline search also spans a heterogeneous target set with variable quarter coverage \citep{Twicken_2016}, so 2.51 should be regarded as an approximate (and plausibly upper-limit) yield scaling.

\begin{figure*}
	\includegraphics[width=2\columnwidth]{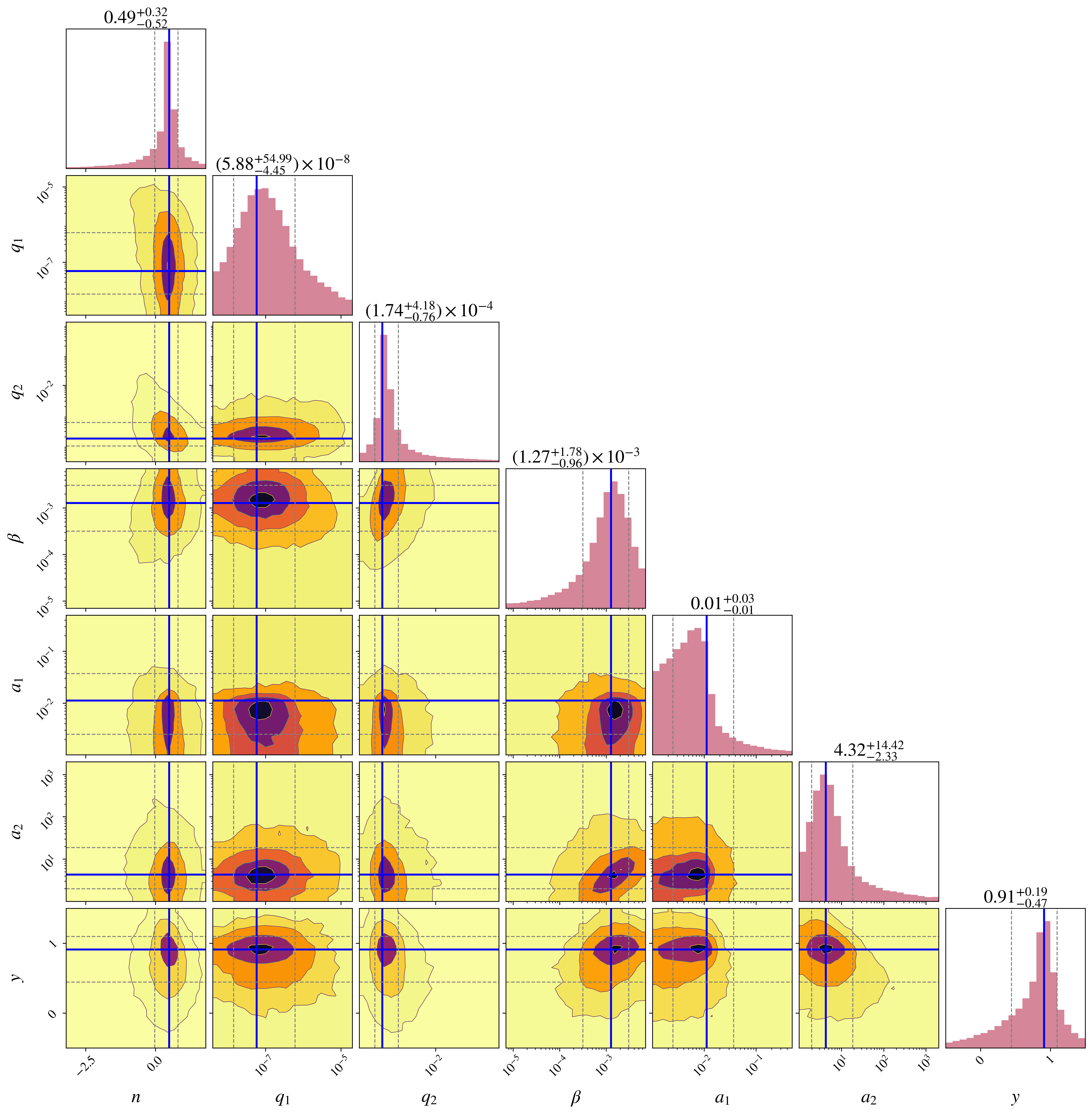}
    \caption{Cornerplot showing 1-D marginals (diagonals) and 2-D joint posteriors (off-diagonals) of samples while fitting for \textit{Kepler} population using differential evolution. The multi-dimensional best-fit value is overplotted in blue colour. The dashed grey lines show the upper-error and the lower-error on the best-fit value.}
\label{fig:cornerplot_de_kepler}
\end{figure*}

\begin{figure*}
	\includegraphics[width=2\columnwidth]{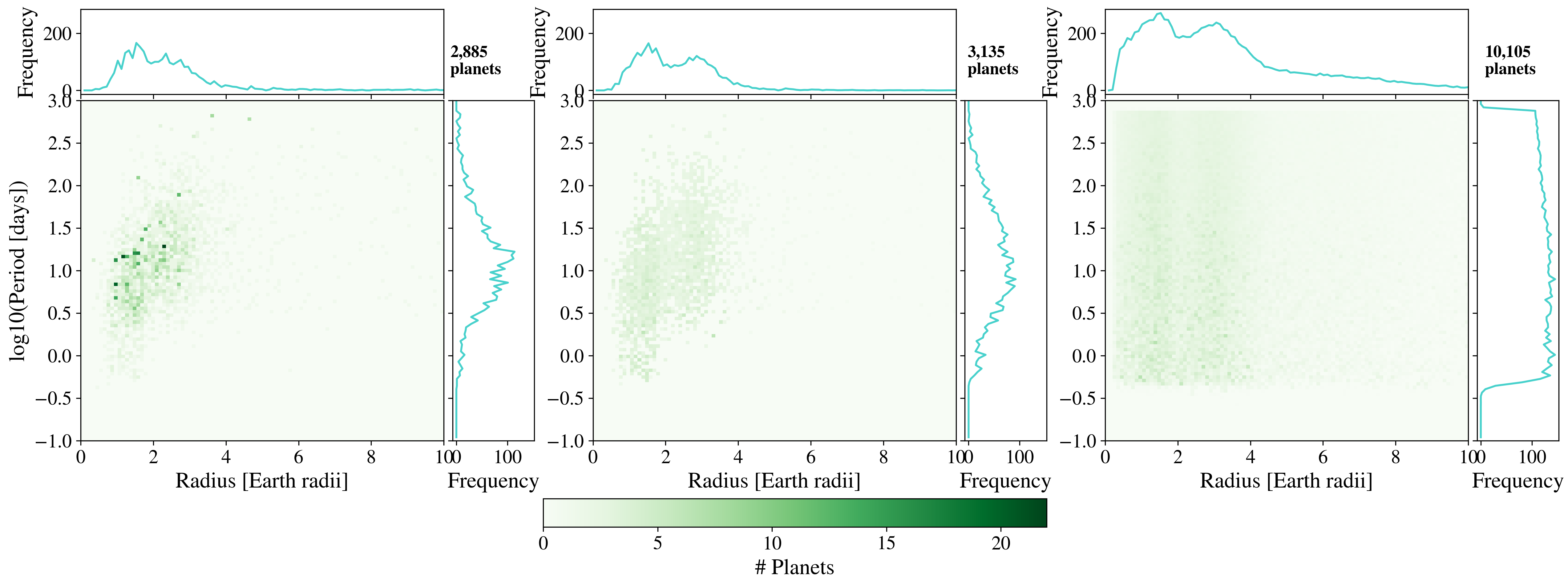}
    \caption{The figure shows the distribution of the projected \textit{Kepler} observations (KeplerPORTs-corrected counts per bin) on the left and the corresponding recovered planets using the 7-parameter model on the middle and right. In the middle panel, we only plot the bins where the projected distribution had non-zero planets. In the rightmost panel, we plot all
the bins from the recovered population. The colourmap shows the number of planets on the period-radius axes. The axes are accompanied by 1-D histograms
of radius and period separately. In the top right corner of each 2-D histogram, the total number of the total planets is written.}
\label{fig:DE_kepler}
\end{figure*}

We use the differential evolution (DE) retrieval to obtain our best-fit parameters, as shown in Table \ref{tab:kepler-test}. The corresponding posterior distributions are shown in Figure~\ref{fig:cornerplot_de_kepler}, and we show the histograms of exoplanet population generated using these values in Figure \ref{fig:DE_kepler}. The recovered planet distribution from the 7-parameter DE model successfully reproduces some key features of the underlying distribution inferred from the \textit{Kepler} population, including the Fulton gap \citep{Fulton_2017} near 
2 R$_{\oplus}$, and the prominent peak in the semi-major axis distribution around $P \sim 10$ days. These agreements indicate that the model captures the dominant demographic trends. However, when the full predicted distribution considering all the bins is examined, the model also generates a significant population of larger planets. Their absence in the observed sample suggests serious deficiencies in this model and also in others that we examine in this paper. 

To mitigate the shortcomings in the 7-parameter model,  we also consider a 9-parameter model and an 11-parameter model. In the 7-parameter model, both $q$ and $a$ are described by single power-law distributions. While in the 9-parameter case, $q$ is allowed to have a broken power law distribution, while $a$ follows a single power law. Finally, in the 11-parameter model, both $q$ and $a$ have broken power-law distributions. The motivation for adopting a broken power-law in semi-major axis arises from the discrepancy between the model and observations: the recovered population exhibits a relatively flat one-dimensional 
$a$-distribution when all bins are considered, whereas the \textit{Kepler} data show a pronounced peak, indicating a possible change in slope across different orbital regimes.

\begin{figure*}
	\includegraphics[width=2\columnwidth]{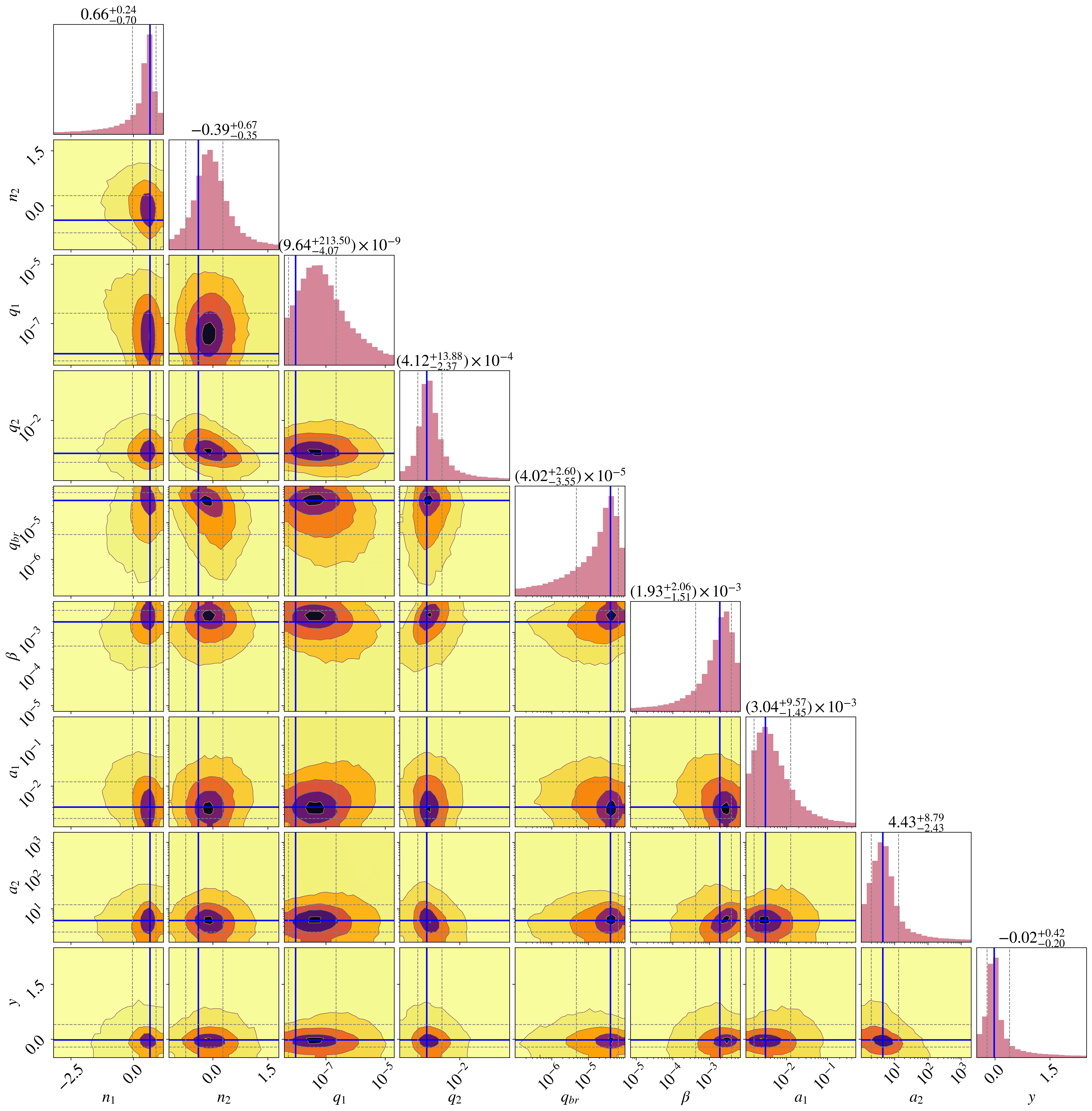}
    \caption{Cornerplot showing 1-D marginals (diagonals) and 2-D joint posteriors (off-diagonals) of samples while fitting for \textit{Kepler} population on 9-parameter model as defined in Section \ref{sec:kepler-params}, using differential evolution. The multi-dimensional best-fit value is overplotted in blue colour. The dashed grey lines show the upper-error and the lower-error on the best-fit value.}
\label{fig:cornerplot_de_kepler_9}
\end{figure*}

\begin{figure*}
	\includegraphics[width=2\columnwidth]{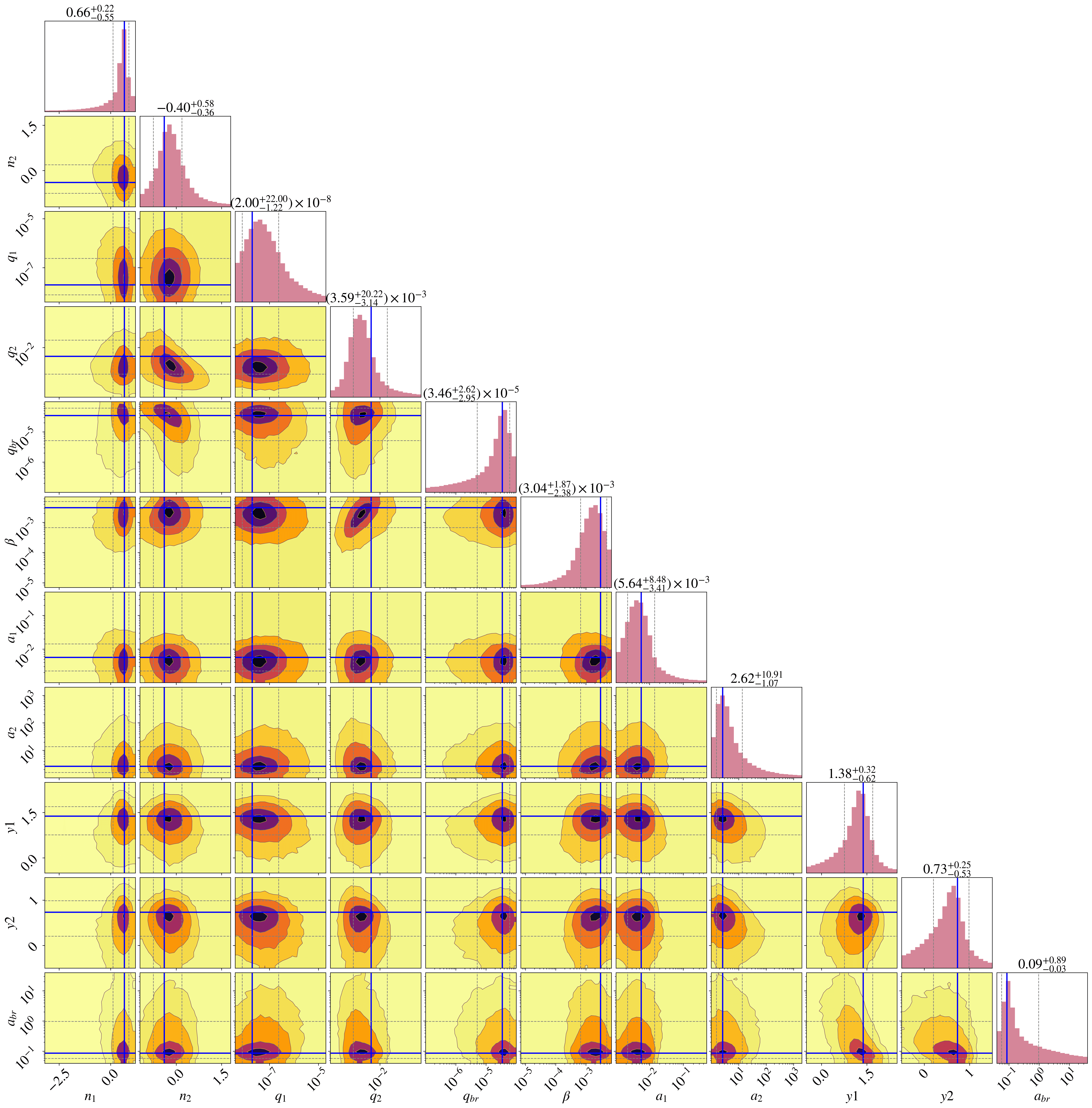}
    \caption{Cornerplot showing 1-D marginals (diagonals) and 2-D joint posteriors (off-diagonals) of samples while fitting for \textit{Kepler} population on 11-parameter model as defined in Section \ref{sec:kepler-params}, using differential evolution. The multi-dimensional best-fit value is overplotted in blue colour. The dashed grey lines show the upper-error and the lower-error on the best-fit value.}
\label{fig:cornerplot_de_kepler_11}
\end{figure*}

\begin{figure*}
	\includegraphics[width=2\columnwidth]{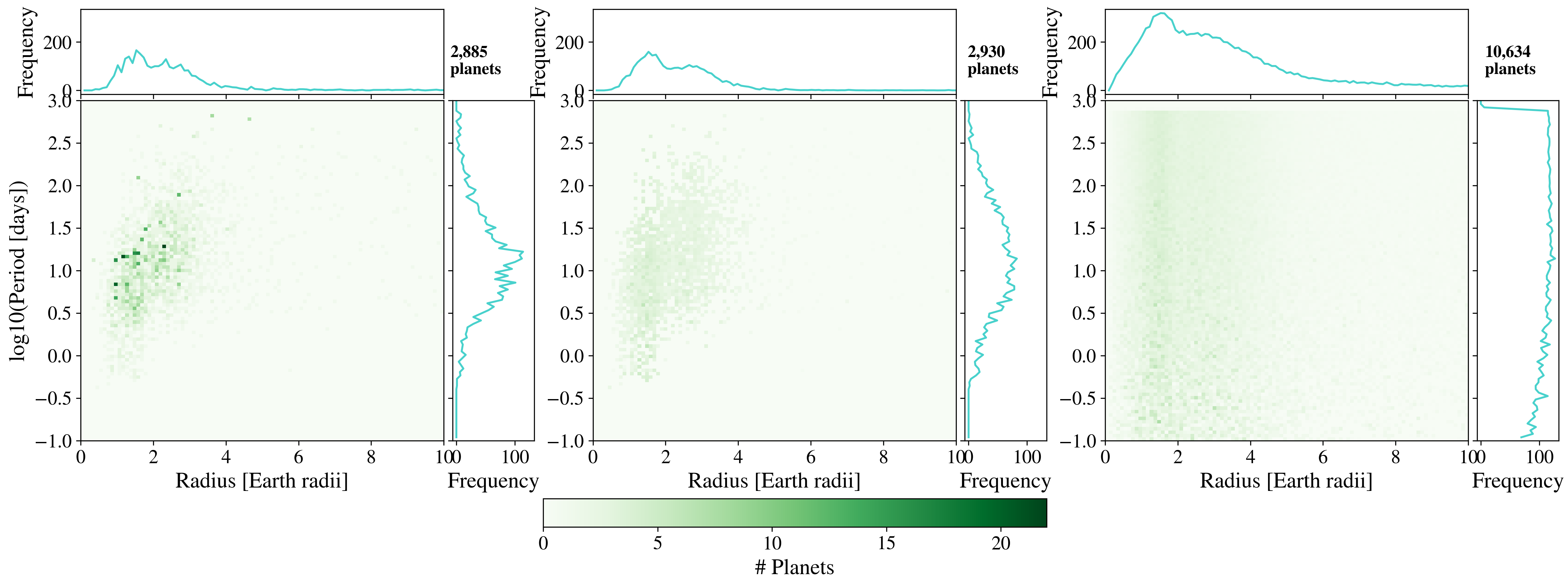}
    \caption{The figure shows the distribution of the projected \textit{Kepler} observations (KeplerPORTs-corrected counts per bin) on the left and the corresponding recovered planets using the 9-parameter model on the middle and right. In the middle panel, we only plot the bins where the projected distribution had non-zero planets. In the rightmost panel, we plot all
the bins from the recovered population. The colourmap shows the number of planets on the period-radius axes. The axes are accompanied by 1-D histograms
of radius and period separately. In the top right corner of each 2-D histogram, the total number of the total planets is written.}
\label{fig:DE_kepler9}
\end{figure*}

\begin{figure*}
	\includegraphics[width=2\columnwidth]{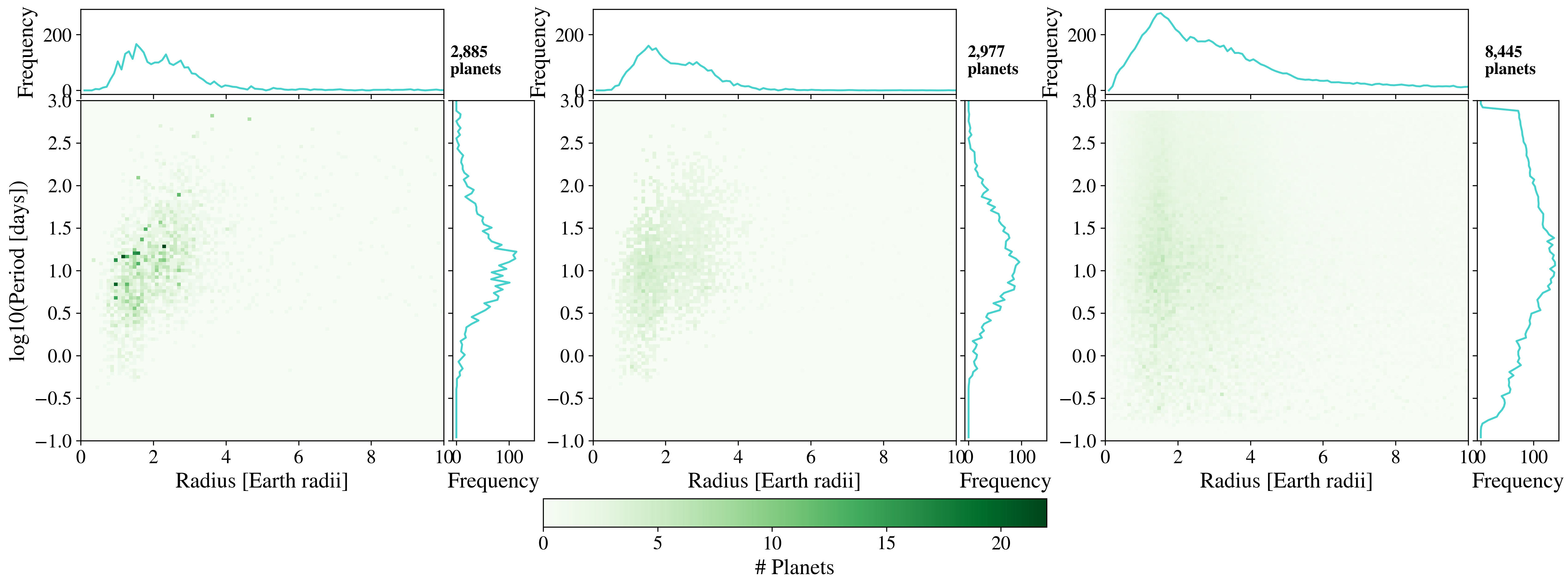}
    \caption{The figure shows the distribution of the projected \textit{Kepler} observations (KeplerPORTs-corrected counts per bin) on the left and the corresponding recovered planets using the 11-parameter model in the middle and right. In the middle panel, we only plot the bins where the projected distribution had non-zero planets. In the rightmost panel, we plot all
the bins from the recovered population. The colourmap shows the number of planets on the period-radius axes. The axes are accompanied by 1-D histograms
of radius and period separately. In the top right corner of each 2-D histogram, the total number of the total planets is written.}
\label{fig:DE_kepler11}
\end{figure*}

The priors and results for the 9-parameter and 11-parameter models are provided in Table \ref{tab:kepler-test}, with the corresponding corner plots shown in Figures \ref{fig:cornerplot_de_kepler_9} and \ref{fig:cornerplot_de_kepler_11}. A comparison of exoplanet population is shown in Figures \ref{fig:DE_kepler9} and \ref{fig:DE_kepler11}. The 9-parameter model shows no improvements over the 7-parameter model, while the 11-parameter model shows a better agreement in the comparability of semi-major axis histograms.

\begin{table*}
    \centering
    \caption[]{The table shows the recovered parameters for \textit{Kepler} observations when using 7-parameter, 9-parameter, and 11-parameter model fitted using differential evolution method.  The priors for the parameters are uniform for $n_1$, $n_2$, $y_1$, and $y_2$, while log-uniform for the others. }
    \label{tab:kepler-test}
    \centering
    \begin{tabular}{lllll}
    \hline
    Parameter & Prior & 7-param & 9-param & 11-param\\[.35em]
    \hline
    $n_1$ ($n$) & [-3.2, 1.2] & $0.49^{+0.32}_{-0.52}$ & $0.66^{+0.24}_{-0.70}$ & $0.66^{+0.22}_{-0.55}$  \\ [0.35em]
$n_2$ & [-1.2, 1.8] & - & $-0.39^{+0.67}_{-0.35}$ & $-0.40^{+0.58}_{-0.36}$  \\ [0.35em]
$q_1$ & [4e-09, 2e-05] & $(5.88^{+54.99}_{-4.45}) \times 10^{-8}$ & $(9.64^{+213.50}_{-4.07}) \times 10^{-9}$ & $(2.00^{+22.00}_{-1.22}) \times 10^{-8}$  \\ [0.35em]
$q_2$ & [3e-05, 1.23] & $(1.74^{+4.18}_{-0.76}) \times 10^{-4}$ & $(4.12^{+13.88}_{-2.37}) \times 10^{-4}$ & $(3.59^{+20.22}_{-3.14}) \times 10^{-3}$  \\ [0.35em]
$q_{br}$ & [1e-07, 0.0001] & - & $(4.02^{+2.60}_{-3.55}) \times 10^{-5}$ & $(3.46^{+2.62}_{-2.95}) \times 10^{-5}$  \\ [0.35em]
$\beta$ & [7e-06, 0.007] & $(1.27^{+1.78}_{-0.96}) \times 10^{-3}$ & $(1.93^{+2.06}_{-1.51}) \times 10^{-3}$ & $(3.04^{+1.87}_{-2.38}) \times 10^{-3}$  \\ [0.35em]
$a_1$ & [0.001, 0.5] & $0.01^{+0.03}_{-0.01}$ & $(3.04^{+9.57}_{-1.45}) \times 10^{-3}$ & $(5.64^{+8.48}_{-3.41}) \times 10^{-3}$  \\ [0.35em]
$a_2$ & [1, 2000] & $4.32^{+14.42}_{-2.33}$ & $4.43^{+8.79}_{-2.43}$ & $2.62^{+10.91}_{-1.07}$  \\ [0.35em]
$y_1$ ($y$) & [-0.5, 2.5] & $0.91^{+0.19}_{-0.47}$ & $-0.02^{+0.42}_{-0.20}$ & $1.38^{+0.32}_{-0.62}$  \\ [0.35em]
$y_2$ & [-0.5, 1.5] & - & - & $0.73^{+0.25}_{-0.53}$  \\ [0.35em]
$a_{br}$ & [0.04, 40] & - & - & $0.09^{+0.89}_{-0.03}$  \\ [0.35em]
Runtime (s) & & 5,899 & 16,305 & 25,755\\
 \hline
    \end{tabular}
\end{table*}

We calculate the Bayesian Information Criterion (BIC) \citep{1978AnSta...6..461S, astropy:2022} to compare the 7, 9, and 11 parameter models. There were 1261 bins which had non-zero planets in \textit{Kepler} observation. This value is used as the number of samples for BIC calculation. The BIC came out to be 4830.98, 4734.68, and 4643.19, for 7, 9, and 11 parameter models, respectively. This shows that the 11-parameter model is preferred.

\subsection{Modelling limitations} \label{sec:limits}

Upon observing the unmasked 2D histograms of predicted \textit{Kepler} exoplanet distribution from the 7,9, and 11 parameter model, we notice that the underlying model predicts a substantial number of planets with  $P=10$ days and $R_{\mathrm{p}}>2 R_{\oplus}$. Such planets should be easily detected by \textit{Kepler} for many hosts, so this points to a clear deficiency of the model in not accounting for possible correlations between $a$ and $q$, as may arise due to Hill instability, for example. 

\begin{figure}
	\includegraphics[width=\columnwidth]{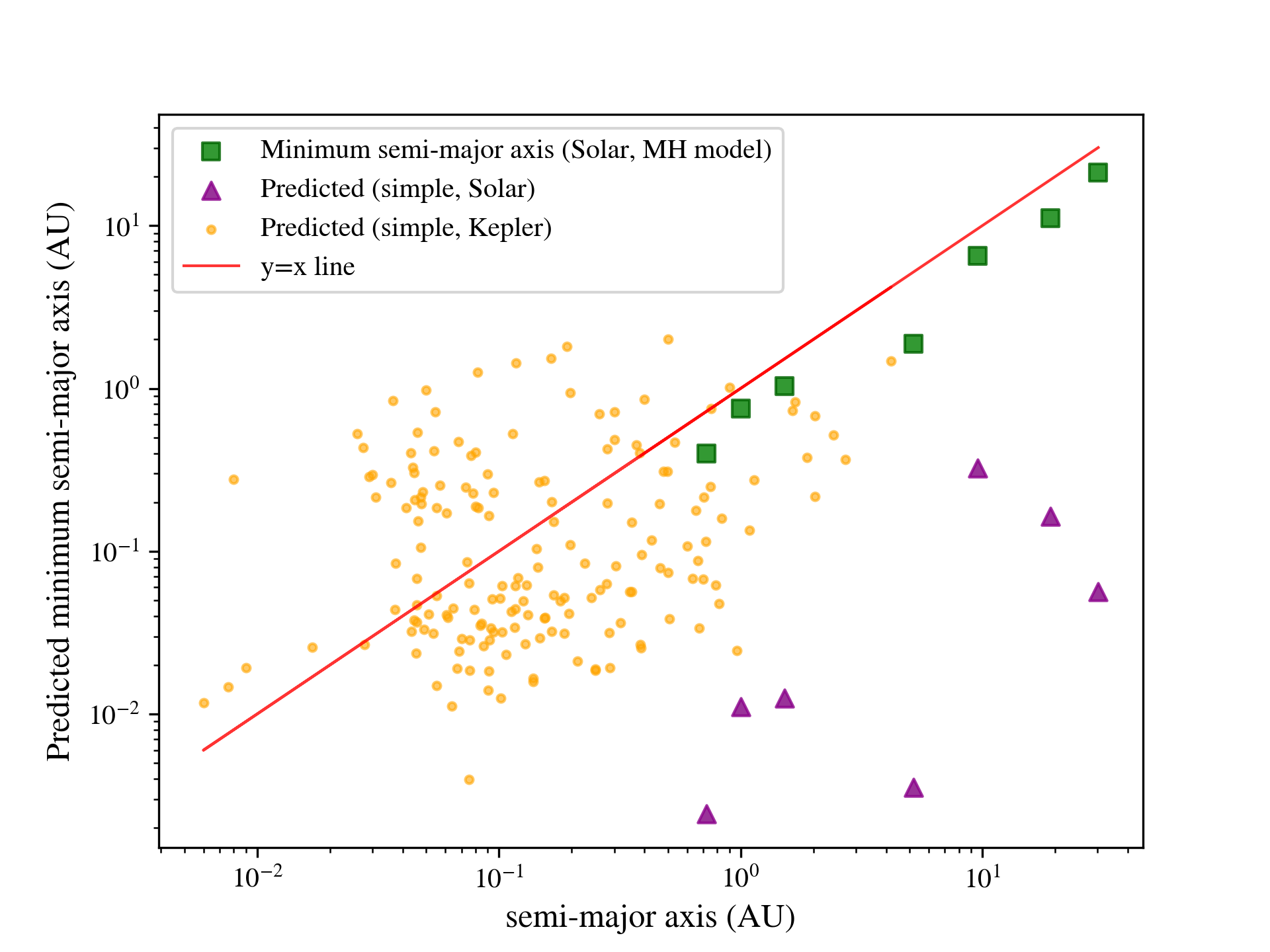}
    \caption{The figure shows the distribution of predicted $a_{min}$ vs observed $a$ for all \textit{Kepler} exoplanets in orange dots, and Solar system planets in purple triangles, calculated using Equation \eqref{eq:mh_simple}. The green squares show the $a_{min}$ calculated using Equation \eqref{eq:mh} for Solar system planets. The red line shows the $y=x$ line. Ideally, we would want all the orange dots to lie close to, and below the red line.}
\label{fig:mh_comparison}
\end{figure}

\begin{figure*}
	\includegraphics[width=2\columnwidth]{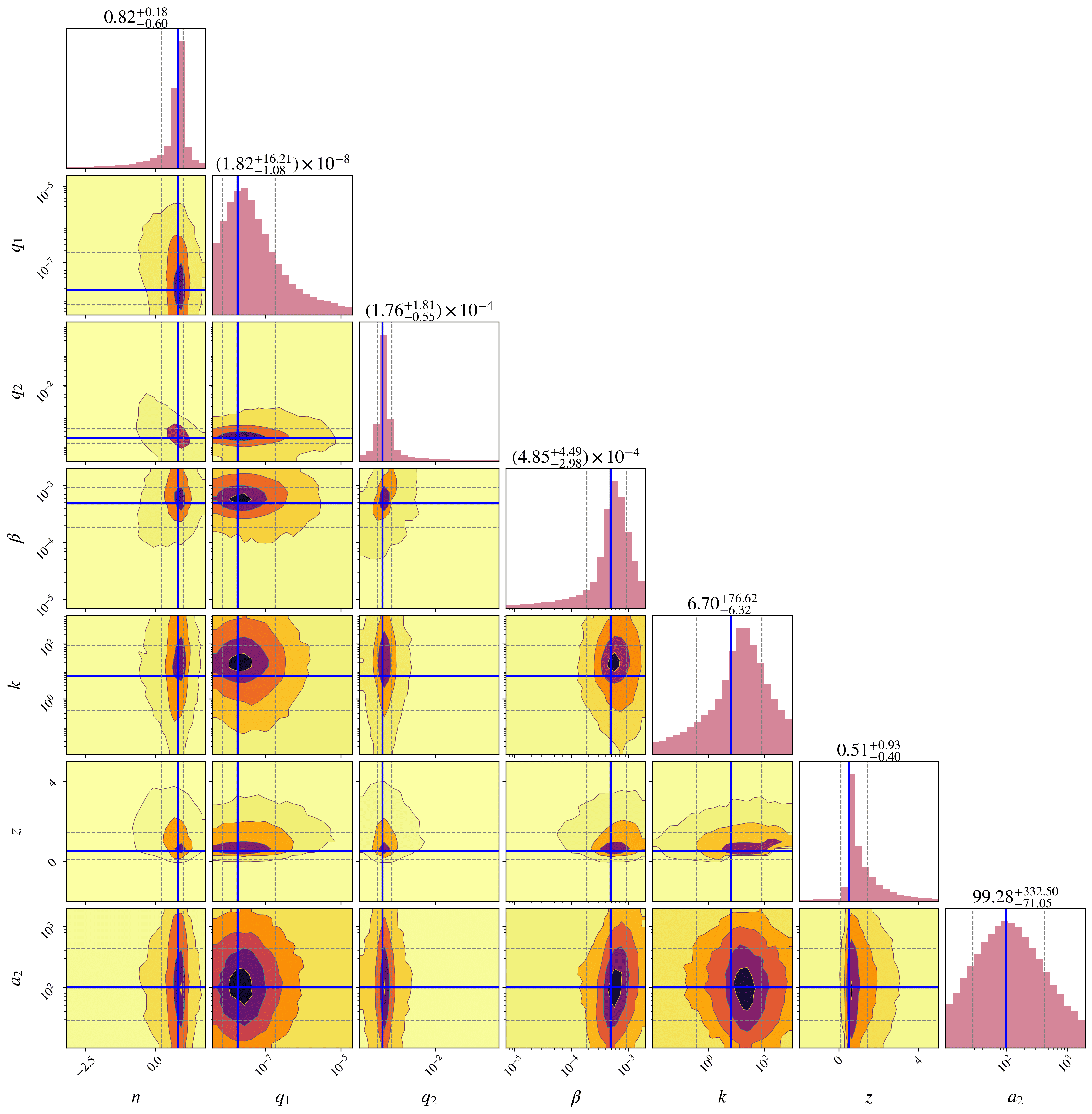}
    \caption{Cornerplot showing 1-D marginals (diagonals) and 2-D joint posteriors (off-diagonals) of samples while fitting for \textit{Kepler} population on 7-parameter CPL model using differential evolution. The multi-dimensional best-fit value is overplotted in blue colour. The dashed grey lines show the upper-error and the lower-error on the best-fit value.}
\label{fig:cornerplot_pl}
\end{figure*}

\begin{figure*}
	\includegraphics[width=2\columnwidth]{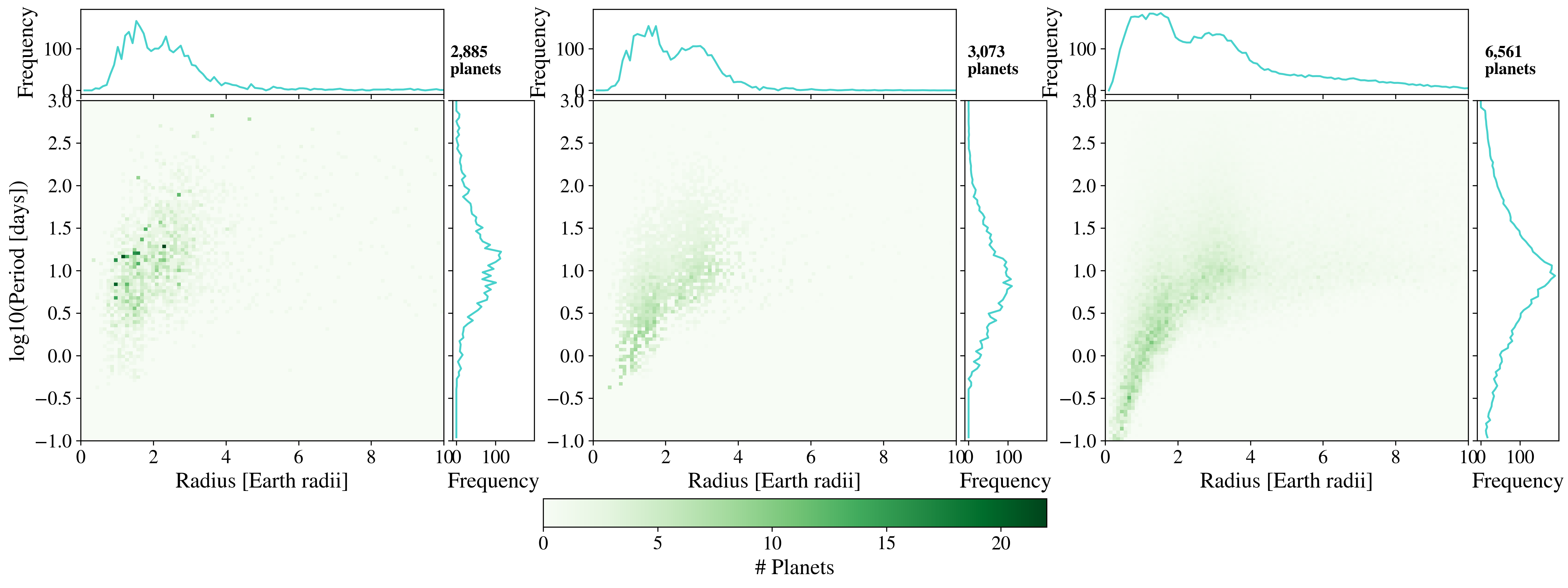}
    \caption{The figure shows the distribution of the projected \textit{Kepler} observations on the left and the corresponding recovered planets using the CPL model (Section \ref{sec:limits}) on the middle and right. In the middle panel, we only plot the bins where the projected distribution had non-zero planets. In the rightmost panel, we plot all
the bins from the recovered population. The colourmap shows the number of planets on the period-radius axes. The axes are accompanied by 1-D histograms
of radius and period separately. In the top right corner of each 2-D histogram, the total number of the total planets is written.}
\label{fig:compare_histogram_pl}
\end{figure*}

For Hill stability, we have \citep{OBERTAS201752},
\begin{equation}
    a_2 > a_1(1+2\times 3^{1/6}\times(q_1+q_2)^{1/3}),
    \label{eq:mh}
\end{equation}
\noindent where the subscripts $1$ and $2$ denote the inner and outer planet. We consider a simplified single planet proxy of this inequality, 
\begin{equation}
    a > k \times q_1^z,
    \label{eq:mh-single}
\end{equation}
\noindent or,
\begin{equation}
    a_{min} = k \times q_1^z,
    \label{eq:mh_simple}
\end{equation}
\noindent where $k$ and $z$ are parameters to fit, and 
\begin{equation}
    a \in \text{LogUniform}(a_{\text{min}}, a_{2}).
    \label{pl-semimaj-sampling}
\end{equation}
Using this $a-q$ correlated power-law (CPL) in our 7-parameter model from \ref{sec:diff-evo}, instead of fitting for $a_1$, we fit for $k$ and $z$. To keep the model simpler, we consider the semi-major distribution to be log-uniform and remove parameter $y$. Subsequently, we find the best-fit values for $k$ and $z$. The BIC for this retrieval was 5789.19. The resulting distributions are shown in Figure \ref{fig:mh_comparison}, and the corresponding best-fit parameters are listed in Table \ref{tab:kepler-test-pl}. The corresponding posterior distribution is shown in Figure \ref{fig:cornerplot_pl}, with comparison histogram shown in Figure \ref{fig:compare_histogram_pl}. The distribution of recovered planets using this model shows improved agreement when comparing the 1D histograms, even after considering all the bins. The existence of a few planets with an orbital period $\sim$10 days and $\geq 4$ R$_{\bigoplus}$ suggests that even this model requires additional physics to suppress larger planets at low periods. Additionally, we neglect stellar multiplicity and treat each BGM catalogue entry as an isolated host, although multiplicity is common in the field \citep{2013ARA&A..51..269D,Raghavan_2010}. This approximation is adequate for the present Kepler-like transit validation, whose primary aim is to test TAED retrieval performance under controlled assumptions. However, ignoring unresolved companions can bias inferred host properties and survey selection (e.g.\ transit-depth dilution) The purpose of this paper is to present the TAED framework; we leave the exploration of better-fitting and more complex models to future studies.  

\begin{table}
    \centering
    \caption[]{The table shows the recovered parameters for \textit{Kepler} observations when using CPL model which incorporates correlated $a$ and $q$ power-law relations, following equation \ref{eq:mh_simple}. The priors for the parameters are uniform for $n$ and $z$, while log-uniform for the others. }
    \label{tab:kepler-test-pl}
    \centering
    \begin{tabular}{lll}
    \hline
    Parameter & Prior & CPL\\[.35em]
    \hline
    $n$ & [-3.2, 1.8] & $0.82^{+0.18}_{-0.60}$  \\ [0.35em]
$q_1$ & [4e-09, 2e-05] & $(1.82^{+16.21}_{-1.08}) \times 10^{-8}$  \\ [0.35em]
$q_2$ & [3e-05, 1.23] & $(1.76^{+1.81}_{-0.55}) \times 10^{-4}$  \\ [0.35em]
$\beta$ & [7e-06, 0.002] & $(4.85^{+4.49}_{-2.98}) \times 10^{-4}$  \\ [0.35em]
$k$ & [0.01, 1000] & $6.70^{+76.62}_{-6.32}$  \\ [0.35em]
$z$ & [-2, 5] & $0.51^{+0.93}_{-0.40}$  \\ [0.35em]
$a_2$ & [10, 2000] & $99.28^{+332.50}_{-71.05}$  \\ [0.35em]
Runtime (s) & & 4,076 \\
 \hline
    \end{tabular}
\end{table}

\subsection{Demographic sub-structure and catalogue reliability}
\label{sec:taed_limitations}
Although the CPL modification suppresses part of the short-period, large-radius excess (Figure \ref{fig:compare_histogram_pl}), the present TAED parameterisation remains a broad description of the underlying distribution. Such a model can capture smooth trends and gross yield normalisation, but it is not designed to reproduce sharp boundaries, localised overdensities, or period-dependent morphology. This is relevant for close-in sub-structure such as the Neptunian desert \citep{neptune_desert}, any period dependence of the radius valley \citep{10.1093/mnras/sty1783}, and the hot-Jupiter pile-up (and potentially a Neptunian ridge) \citep{2024A&A...689A.250C,2025arXiv251026882W}. In that sense, the Hill-stability-motivated correlation should be viewed as a partial structural constraint rather than a dominant driver of close-in demographic signatures, which are more plausibly linked to atmospheric loss and tidal evolution \citep{10.1093/mnras/sty1760}. Additional limitations include the assumption of circular orbits (affecting transit probability and the inferred \(a\)-distribution) and the mapping from \(q\) to \(R_p\), which has intrinsic scatter in mass--radius relations \citep{2026MNRAS.546ag088E}.

The retrieval also implicitly assumes that the completeness-corrected counts represent bona fide planets; residual false positives or false alarms would distort \(N_{\mathrm{obs},\epsilon}\) in a non-uniform way across \((P,R_p)\) \citep{Morton_2016}, and could bias both the inferred normalisation (e.g.\ \(\beta\)) and slopes. It may therefore be worth considering whether bin-wise reliability weights or candidate-level false-positive probabilities \citep{2018ApJS..235...38T} could be incorporated, so that regions of lower reliability contribute less strongly to the likelihood.

These considerations are mentioned for completeness, but we do not attempt to model reliability explicitly here.

\section{Conclusions} \label{sec:conc}

The NASA \textit{Roman} Galactic Bulge Time Domain Survey is a five-year high-cadence survey to be undertaken by \textit{Roman} from early 2027. It will hugely expand both the number and locations of known exoplanets. Up to 200,000 exoplanets are expected to be discovered through the transit and microlensing detection techniques, and these will be located towards, in, and beyond the Galactic bulge. \textit{Roman} will access large numbers of hot and cool exoplanets around a representative mix of stellar hosts for the first time. \textit{Roman}'s sample will present a treasure trove for studies of Galactic-wide exoplanet demography and will provide stringent new tests for planet formation models. 

With this opportunity comes a new challenge: to develop exoplanet demographic analysis methods that can combine the full set of information available from large microlensing and transit samples, given the differences in properties and Galactic locations of the planets and their hosts within each sample. To address this challenge, we introduce in this paper a \textit{technique-agnostic exoplanet demography} (TAED) framework that is able to make consistent forecasts for multiple detection methods that are based on spatio-kinematic properties. This includes transits, microlensing, radial velocity and astrometric detection methods. 

This paper presents the design and first test of the TAED framework, focusing on simulations of a \textit{Kepler}-like transit observable. We have examined the accuracy and scalability of the recovery of exoplanet demographic parameters from simulated transit datasets where the truth is known. We have validated the TAED framework and have demonstrated a workable retrieval method, based on differential evolution, which can recover the underlying exoplanet population and can scale to large datasets of the kind that \textit{Roman} is expected to deliver. Differential evolution was found to be preferred over several other retrieval methods we investigated, including nested sampling and machine learning, based on its combination of speed and accuracy of recovery. 

For our initial test of the TAED framework, we considered several fairly simple toy exoplanet demographic models with 7 to 11 free parameters. In our tests, the TAED framework was successful in retrieving key population features and parameters for our injected distributions. We also applied the TAED retrieval to the \textit{Kepler} DR25 dataset itself, even though the assumed models have known simplicities and limitations. The recovered models successfully reproduced major known demographic features, including the Fulton gap, and the peak in orbital period distribution around 10 days. However, due to their simplicity, the models typically over-predicted exoplanets in regions of parameter space where \textit{Kepler} found few or none. This does not point to a deficiency in the TAED framework itself. Rather, it highlights how the framework can be used to evolve towards more complex exoplanet models that include additional covariances between planet parameters and between planet and host properties that facilitate a better match to the observations. It was not the aim of this current paper to undertake such a study, though it is clearly a goal for future study.

The logical next step will be to test TAED on simulated microlensing samples, and subsequently on combined fully-scaled transit and microlensing samples, based on the characteristics of the expected \textit{Roman} survey sensitivity. The prospect of deploying TAED on combined samples is particularly exciting. Microlensing and transit methods have different sensitivity to planet size/mass, orbital period/radius, orbit eccentricity and, when considering multi-planet systems, mutual inclination. Using the TAED framework, we expect demographic parameter retrieval from joint microlensing and transit samples to provide tighter constraints on some of these distributions than would arise from studies of the separate samples. This strongly motivates a joint approach to \textit{Roman} microlensing and transit datasets for demographic studies.

Lastly, given that transit and microlensing samples represent hot and cold exoplanets that typically straddle either side of the stellar habitable zone, a joint demographic analysis of the \textit{Roman} transit and microlensing samples likely offers one of the most reliable calibrations of the total Galactic occurrence of habitable zone exoplanets. This would represent the first true measurement of the third term in the Drake Equation. 
%%%%%%%%%%%%%%%%%%%%%%%%%%%%%%%%%%%%%%%%%%%%%%%%%%%%%%%%%%%%%%
\begin{acknowledgements}

\end{acknowledgements}
This paper includes data collected by the \textit{Kepler} mission, which is funded by the NASA Discovery Programme. AP is supported by a PhD studentship from the United Kingdom's Science and Technology Facilities Council (STFC). Language refinement in parts of this manuscript was supported by Copilot.

% Please note that we have included the references below in
% order to compile the document, but we ask you to:

%\begin{thebibliography}{}
\bibliographystyle{aa}
\bibliography{example}

@ARTICLE{2021JOSS....6.3001B,
       author = {{Buchner}, Johannes},
        title = "{UltraNest - a robust, general purpose Bayesian inference engine}",
      journal = {The Journal of Open Source Software},
         year = 2021,
        month = apr,
       volume = {6},
       number = {60},
          eid = {3001},
        pages = {3001},
          doi = {10.21105/joss.03001},
archivePrefix = {arXiv},
       eprint = {2101.09604},
 primaryClass = {stat.CO},
       adsurl = {https://ui.adsabs.harvard.edu/abs/2021JOSS....6.3001B}
}

@ARTICLE{2026MNRAS.546ag088E,
       author = {{Edmondson}, Kathryn and {Kerins}, Eamonn},
        title = "{A planet─host ratio relation to synthesize microlensing and transiting exoplanet demography from Roman}",
      journal = {\mnras},
         year = 2026,
        month = feb,
       volume = {546},
       number = {2},
          eid = {stag088},
        pages = {stag088},
          doi = {10.1093/mnras/stag088},
archivePrefix = {arXiv},
       eprint = {2506.24004},
 primaryClass = {astro-ph.EP},
       adsurl = {https://ui.adsabs.harvard.edu/abs/2026MNRAS.546ag088E}
}

@article{Robin2003AWay,
    title = {{A synthetic view on structure and evolution of the Milky Way}},
    year = {2003},
    journal = {Astronomy and Astrophysics},
    fjournal = {Astronomy and Astrophysics},
    author = {Robin, A. C. and Reyl{\'{e}}, C. and Derri{\`{e}}re, S. and Picaud, S.},
    number = {2},
    month = {10},
    pages = {523--540},
    volume = {409},
    doi = {10.1051/0004-6361:20031117},
    issn = {00046361},
    arxivId = {astro-ph/0401052}
}

@article{Pascucci2018AStars,
    title = {{A Universal Break in the Planet-to-star Mass-ratio Function of Kepler MKG Stars}},
    year = {2018},
    journal = {The Astrophysical Journal Letters},
    fjournal = {The Astrophysical Journal Letters},
    author = {Pascucci, Ilaria and Mulders, Gijs D. and Gould, Andrew and Fernandes, Rachel},
    number = {2},
    month = {3},
    pages = {L28},
    volume = {856},
    publisher = {IOP Publishing},
    url = {https://iopscience.iop.org/article/10.3847/2041-8213/aab6ac https://iopscience.iop.org/article/10.3847/2041-8213/aab6ac/meta},
    doi = {10.3847/2041-8213/AAB6AC},
    issn = {2041-8205},
    arxivId = {1803.00777}
}

@article{Jordi2006EmpiricalSystems,
    title = {{Empirical color transformations between SDSS photometry and other photometric systems}},
    year = {2006},
    journal = {Astronomy and Astrophysics},
    fjournal = {Astronomy and Astrophysics},
    author = {Jordi, K. and Grebel, E. K. and Ammon, K.},
    number = {1},
    month = {12},
    pages = {339--347},
    volume = {460},
    url = {https://ui.adsabs.harvard.edu/abs/2006A&A...460..339J/abstract},
    doi = {10.1051/0004-6361:20066082},
    issn = {00046361},
    arxivId = {astro-ph/0609121}
}

@article{Hsu2018ImprovingComputation,
    title = {{Improving the Accuracy of Planet Occurrence Rates from Kepler Using Approximate Bayesian Computation}},
    year = {2018},
    journal = {The Astronomical Journal},
    fjournal = {The Astronomical Journal},
    author = {Hsu, Danley C. and Ford, Eric B. and Ragozzine, Darin and Morehead, Robert C.},
    number = {5},
    month = {4},
    pages = {205},
    volume = {155},
    publisher = {IOP Publishing},
    url = {https://iopscience.iop.org/article/10.3847/1538-3881/aab9a8 https://iopscience.iop.org/article/10.3847/1538-3881/aab9a8/meta},
    doi = {10.3847/1538-3881/AAB9A8},
    issn = {1538-3881},
    arxivId = {1803.10787}
}

@article{Brown2011KEPLERCLASSIFICATION,
    title = {{KEPLER INPUT CATALOG: PHOTOMETRIC CALIBRATION AND STELLAR CLASSIFICATION}},
    year = {2011},
    journal = {The Astronomical Journal},
    fjournal = {The Astronomical Journal},
    author = {Brown, Timothy M. and Latham, David W. and Everett, Mark E. and Esquerdo, Gilbert A.},
    number = {4},
    month = {9},
    pages = {112},
    volume = {142},
    publisher = {IOP Publishing},
    url = {https://iopscience.iop.org/article/10.1088/0004-6256/142/4/112 https://iopscience.iop.org/article/10.1088/0004-6256/142/4/112/meta},
    isbn = {5.14115.8381},
    doi = {10.1088/0004-6256/142/4/112},
    issn = {1538-3881},
    arxivId = {1102.0342}
}

@article{Marshall2006ModellingDimensions,
    title = {{Modelling the Galactic interstellar extinction distribution in three dimensions}},
    year = {2006},
    journal = {Astronomy and Astrophysics},
    fjournal = {Astronomy and Astrophysics},
    author = {Marshall, D. J. and Robin, A. C. and Reyl{\'{e}}, C. and Schultheis, M. and Picaud, S.},
    number = {2},
    month = {7},
    pages = {635--651},
    volume = {453},
    doi = {10.1051/0004-6361:20053842},
    issn = {00046361}
}

@article{Burke2015TERRESTRIALSAMPLE,
    title = {{TERRESTRIAL PLANET OCCURRENCE RATES FOR THE KEPLER GK DWARF SAMPLE}},
    year = {2015},
    journal = {The Astrophysical Journal},
    fjournal = {The Astrophysical Journal},
    author = {Burke, Christopher J. and Christiansen, Jessie L. and Mullally, F. and Seader, Shawn and Huber, Daniel and Rowe, Jason F. and Coughlin, Jeffrey L. and Thompson, Susan E. and Catanzarite, Joseph and Clarke, Bruce D. and Morton, Timothy D. and Caldwell, Douglas A. and Bryson, Stephen T. and Haas, Michael R. and Batalha, Natalie M. and Jenkins, Jon M. and Tenenbaum, Peter and Twicken, Joseph D. and Li, Jie and Quintana, Elisa and Barclay, Thomas and Henze, Christopher E. and Borucki, William J. and Howell, Steve B. and Still, Martin},
    number = {1},
    month = {8},
    pages = {8},
    volume = {809},
    publisher = {IOP Publishing},
    url = {https://iopscience.iop.org/article/10.1088/0004-637X/809/1/8 https://iopscience.iop.org/article/10.1088/0004-637X/809/1/8/meta},
    doi = {10.1088/0004-637X/809/1/8},
    issn = {0004-637X},
    arxivId = {1506.04175}
}

@incollection{Gaudi2021TheExoplanets,
    title = {{The Demographics of Exoplanets}},
    year = {2021},
    booktitle = {ExoFrontiers},
    author = {Gaudi, B. Scott and Meyer, Michael and Christiansen, Jessie},
    month = {10},
    series = {2514-3433},
    type = {Book Chapter},
    pages = {2-1 to 2-21},
    publisher = {IOP Publishing},
    doi = {10.1088/2514-3433/ABFA8FCH2},
    isbn = {978-0-7503-1472-5},
    arxivId = {2011.04703}
}

@article{Mulders2018TheSystems,
    title = {{The Exoplanet Population Observation Simulator. I. The Inner Edges of Planetary Systems}},
    year = {2018},
    journal = {The Astronomical Journal},
    fjournal = {The Astronomical Journal},
    author = {Mulders, Gijs D. and Pascucci, Ilaria and Apai, Dániel and Ciesla, Fred J.},
    number = {1},
    month = {6},
    pages = {24},
    volume = {156},
    publisher = {IOP Publishing},
    url = {https://iopscience.iop.org/article/10.3847/1538-3881/aac5ea https://iopscience.iop.org/article/10.3847/1538-3881/aac5ea/meta},
    doi = {10.3847/1538-3881/AAC5EA},
    issn = {1538-3881},
    arxivId = {1805.08211}
}

@MISC{2017ksci.rept...19B,
       author = {{Burke}, Christopher J. and {Catanzarite}, Joseph},
        title = "{Planet Detection Metrics: Per-Target Detection Contours for Data Release 25}",
 howpublished = {KSCI-19111-002},
         year = 2017,
        month = jun,
          eid = {19},
        pages = {19},
       adsurl = {https://ui.adsabs.harvard.edu/abs/2017ksci.rept...19B}
}

@misc{zhang2017gpuaccelerationlargescaletreeboosting,
      title={GPU-acceleration for Large-scale Tree Boosting}, 
      author={Huan Zhang and Si Si and Cho-Jui Hsieh},
      year={2017},
      eprint={1706.08359},
      archivePrefix={arXiv},
      primaryClass={stat.ML},
      url={https://arxiv.org/abs/1706.08359}, 
}

@ARTICLE{2020MNRAS.493.3132S,
       author = {{Speagle}, Joshua S.},
        title = "{DYNESTY: a dynamic nested sampling package for estimating Bayesian posteriors and evidences}",
      journal = {\mnras},
         year = 2020,
        month = apr,
       volume = {493},
       number = {3},
        pages = {3132-3158},
          doi = {10.1093/mnras/staa278},
archivePrefix = {arXiv},
       eprint = {1904.02180},
 primaryClass = {astro-ph.IM},
       adsurl = {https://ui.adsabs.harvard.edu/abs/2020MNRAS.493.3132S}
}

@ARTICLE{2023arXiv231016733E,
       author = {{Edmondson}, Kathryn and {Norris}, Jordan and {Kerins}, Eamonn},
        title = "{Breaking up with the continuous exoplanet mass-radius relation}",
      journal = {arXiv e-prints},
         year = 2023,
        month = oct,
          eid = {arXiv:2310.16733},
        pages = {arXiv:2310.16733},
          doi = {10.48550/arXiv.2310.16733},
archivePrefix = {arXiv},
       eprint = {2310.16733},
 primaryClass = {astro-ph.EP},
       adsurl = {https://ui.adsabs.harvard.edu/abs/2023arXiv231016733E}
}

@article{Fulton_2017,
doi = {10.3847/1538-3881/aa80eb},
url = {https://dx.doi.org/10.3847/1538-3881/aa80eb},
year = {2017},
month = {aug},
publisher = {The American Astronomical Society},
volume = {154},
number = {3},
pages = {109},
author = {Fulton, Benjamin J. and Petigura, Erik A. and Howard, Andrew W. and Isaacson, Howard and Marcy, Geoffrey W. and Cargile, Phillip A. and Hebb, Leslie and Weiss, Lauren M. and Johnson, John Asher and Morton, Timothy D. and Sinukoff, Evan and Crossfield, Ian J. M. and Hirsch, Lea A.},
title = {The California-Kepler Survey. III. A Gap in the Radius Distribution of Small Planets*},
journal = {The Astronomical Journal}
}

@article{Wilson_2023,
doi = {10.3847/1538-4365/acf3df},
url = {https://dx.doi.org/10.3847/1538-4365/acf3df},
year = {2023},
month = {oct},
publisher = {The American Astronomical Society},
volume = {269},
number = {1},
pages = {5},
author = {Wilson, Robert F. and Barclay, Thomas and Powell, Brian P. and Schlieder, Joshua and Hedges, Christina and Montet, Benjamin T. and Quintana, Elisa and Mcdonald, Iain and Penny, Matthew T. and Espinoza, Néstor and Kerins, Eamonn},
title = {Transiting Exoplanet Yields for the Roman Galactic Bulge Time Domain Survey Predicted from Pixel-level Simulations},
journal = {The Astrophysical Journal Supplement Series}
}

@article{Montet_2017,
doi = {10.1088/1538-3873/aa57fb},
url = {https://dx.doi.org/10.1088/1538-3873/aa57fb},
year = {2017},
month = {feb},
publisher = {The Astronomical Society of the Pacific},
volume = {129},
number = {974},
pages = {044401},
author = {Montet, Benjamin T. and Yee, Jennifer C. and Penny, Matthew T.},
title = {Measuring the Galactic Distribution of Transiting Planets with WFIRST},
journal = {Publications of the Astronomical Society of the Pacific}
}

@article{Perryman_2014,
doi = {10.1088/0004-637X/797/1/14},
url = {https://dx.doi.org/10.1088/0004-637X/797/1/14},
year = {2014},
month = {nov},
publisher = {The American Astronomical Society},
volume = {797},
number = {1},
pages = {14},
author = {Perryman, Michael and Hartman, Joel and Bakos, Gaspar A. and Lindegren, Lennart},
title = {ASTROMETRIC EXOPLANET DETECTION WITH GAIA},
journal = {The Astrophysical Journal}
}

@article{storn_differential_1997,
	title = {Differential {Evolution} – {A} {Simple} and {Efficient} {Heuristic} for global {Optimization} over {Continuous} {Spaces}},
	volume = {11},
	issn = {1573-2916},
	url = {https://doi.org/10.1023/A:1008202821328},
	doi = {10.1023/A:1008202821328},
	number = {4},
	journal = {Journal of Global Optimization},
	author = {Storn, Rainer and Price, Kenneth},
	month = dec,
	year = {1997},
	pages = {341--359},
}

@ARTICLE{2020SciPy-NMeth,
  author  = {Virtanen, Pauli and Gommers, Ralf and Oliphant, Travis E. and
            Haberland, Matt and Reddy, Tyler and Cournapeau, David and
            Burovski, Evgeni and Peterson, Pearu and Weckesser, Warren and
            Bright, Jonathan and {van der Walt}, St{\'e}fan J. and
            Brett, Matthew and Wilson, Joshua and Millman, K. Jarrod and
            Mayorov, Nikolay and Nelson, Andrew R. J. and Jones, Eric and
            Kern, Robert and Larson, Eric and Carey, C J and
            Polat, {\.I}lhan and Feng, Yu and Moore, Eric W. and
            {VanderPlas}, Jake and Laxalde, Denis and Perktold, Josef and
            Cimrman, Robert and Henriksen, Ian and Quintero, E. A. and
            Harris, Charles R. and Archibald, Anne M. and
            Ribeiro, Ant{\^o}nio H. and Pedregosa, Fabian and
            {van Mulbregt}, Paul and {SciPy 1.0 Contributors}},
  title   = {{{SciPy} 1.0: Fundamental Algorithms for Scientific
            Computing in Python}},
  journal = {Nature Methods},
  year    = {2020},
  volume  = {17},
  pages   = {261--272},
  adsurl  = {https://rdcu.be/b08Wh},
  doi     = {10.1038/s41592-019-0686-2},
}

@article{Sob67,
  author = {I. M. Sobol'},
  title = {On the distribution of points in a cube and the approximate evaluation of integrals},
  journal = {Zh. Vychisl. Mat. Mat. Fiz.},
  year = {1967},
  volume = {7},
  number = {4},
  pages = {86--112},
  doi = {10.1016/0041-5553(67)90144-9},
  note = {Translated from {Zh. Vychisl. Mat. Mat. Fiz.} {7}(4) (1967) 784--802}
}

@article{Jeyakumar_2011,
   title={Convergence Analysis of Differential Evolution Variants on Unconstrained Global Optimization Functions},
   volume={2},
   ISSN={0976-2191},
   url={http://dx.doi.org/10.5121/ijaia.2011.2209},
   DOI={10.5121/ijaia.2011.2209},
   number={2},
   journal={International Journal of Artificial Intelligence \&; Applications},
   publisher={Academy and Industry Research Collaboration Center (AIRCC)},
   author={Jeyakumar, G and Shanmugavelayutham, C},
   year={2011},
   month=apr, pages={116–127} }

@article{Qiang2014AUD,
  title={A Unified Differential Evolution Algorithm for Global Optimization},
  author={Ji Qiang and Chad E. Mitchell},
  journal={IEEE Transactions on Evolutionary Computation},
  year={2014},
  url={https://api.semanticscholar.org/CorpusID:55922233}
}

@ARTICLE{astropy:2022,
       author = {{Astropy Collaboration} and {Price-Whelan}, Adrian M. and {Lim}, Pey Lian and {Earl}, Nicholas and {Starkman}, Nathaniel and {Bradley}, Larry and {Shupe}, David L. and {Patil}, Aarya A. and {Corrales}, Lia and {Brasseur}, C.~E. and {N{"o}the}, Maximilian and {Donath}, Axel and {Tollerud}, Erik and {Morris}, Brett M. and {Ginsburg}, Adam and {Vaher}, Eero and {Weaver}, Benjamin A. and {Tocknell}, James and {Jamieson}, William and {van Kerkwijk}, Marten H. and {Robitaille}, Thomas P. and {Merry}, Bruce and {Bachetti}, Matteo and {G{"u}nther}, H. Moritz and {Aldcroft}, Thomas L. and {Alvarado-Montes}, Jaime A. and {Archibald}, Anne M. and {B{'o}di}, Attila and {Bapat}, Shreyas and {Barentsen}, Geert and {Baz{'a}n}, Juanjo and {Biswas}, Manish and {Boquien}, M{'e}d{'e}ric and {Burke}, D.~J. and {Cara}, Daria and {Cara}, Mihai and {Conroy}, Kyle E. and {Conseil}, Simon and {Craig}, Matthew W. and {Cross}, Robert M. and {Cruz}, Kelle L. and {D'Eugenio}, Francesco and {Dencheva}, Nadia and {Devillepoix}, Hadrien A.~R. and {Dietrich}, J{"o}rg P. and {Eigenbrot}, Arthur Davis and {Erben}, Thomas and {Ferreira}, Leonardo and {Foreman-Mackey}, Daniel and {Fox}, Ryan and {Freij}, Nabil and {Garg}, Suyog and {Geda}, Robel and {Glattly}, Lauren and {Gondhalekar}, Yash and {Gordon}, Karl D. and {Grant}, David and {Greenfield}, Perry and {Groener}, Austen M. and {Guest}, Steve and {Gurovich}, Sebastian and {Handberg}, Rasmus and {Hart}, Akeem and {Hatfield-Dodds}, Zac and {Homeier}, Derek and {Hosseinzadeh}, Griffin and {Jenness}, Tim and {Jones}, Craig K. and {Joseph}, Prajwel and {Kalmbach}, J. Bryce and {Karamehmetoglu}, Emir and {Ka{l}uszy{'n}ski}, Miko{l}aj and {Kelley}, Michael S.~P. and {Kern}, Nicholas and {Kerzendorf}, Wolfgang E. and {Koch}, Eric W. and {Kulumani}, Shankar and {Lee}, Antony and {Ly}, Chun and {Ma}, Zhiyuan and {MacBride}, Conor and {Maljaars}, Jakob M. and {Muna}, Demitri and {Murphy}, N.~A. and {Norman}, Henrik and {O'Steen}, Richard and {Oman}, Kyle A. and {Pacifici}, Camilla and {Pascual}, Sergio and {Pascual-Granado}, J. and {Patil}, Rohit R. and {Perren}, Gabriel I. and {Pickering}, Timothy E. and {Rastogi}, Tanuj and {Roulston}, Benjamin R. and {Ryan}, Daniel F. and {Rykoff}, Eli S. and {Sabater}, Jose and {Sakurikar}, Parikshit and {Salgado}, Jes{'u}s and {Sanghi}, Aniket and {Saunders}, Nicholas and {Savchenko}, Volodymyr and {Schwardt}, Ludwig and {Seifert-Eckert}, Michael and {Shih}, Albert Y. and {Jain}, Anany Shrey and {Shukla}, Gyanendra and {Sick}, Jonathan and {Simpson}, Chris and {Singanamalla}, Sudheesh and {Singer}, Leo P. and {Singhal}, Jaladh and {Sinha}, Manodeep and {Sip{H{o}}cz}, Brigitta M. and {Spitler}, Lee R. and {Stansby}, David and {Streicher}, Ole and {{{S}}umak}, Jani and {Swinbank}, John D. and {Taranu}, Dan S. and {Tewary}, Nikita and {Tremblay}, Grant R. and {Val-Borro}, Miguel de and {Van Kooten}, Samuel J. and {Vasovi{'c}}, Zlatan and {Verma}, Shresth and {de Miranda Cardoso}, Jos{'e} Vin{'i}cius and {Williams}, Peter K.~G. and {Wilson}, Tom J. and {Winkel}, Benjamin and {Wood-Vasey}, W.~M. and {Xue}, Rui and {Yoachim}, Peter and {Zhang}, Chen and {Zonca}, Andrea and {Astropy Project Contributors}},
        title = "{The Astropy Project: Sustaining and Growing a Community-oriented Open-source Project and the Latest Major Release (v5.0) of the Core Package}",
      journal = {The Astrophysical Journal},
         year = 2022,
        month = aug,
       volume = {935},
       number = {2},
          eid = {167},
        pages = {167},
          doi = {10.3847/1538-4357/ac7c74},
archivePrefix = {arXiv},
       eprint = {2206.14220},
 primaryClass = {astro-ph.IM},
       adsurl = {https://ui.adsabs.harvard.edu/abs/2022ApJ...935..167A}
}

@article{OBERTAS201752,
title = {The stability of tightly-packed, evenly-spaced systems of Earth-mass planets orbiting a Sun-like star},
journal = {Icarus},
volume = {293},
pages = {52-58},
year = {2017},
issn = {0019-1035},
doi = {https://doi.org/10.1016/j.icarus.2017.04.010},
url = {https://www.sciencedirect.com/science/article/pii/S0019103516302962},
author = {Alysa Obertas and Christa {Van Laerhoven} and Daniel Tamayo}
}

@article{Johnson_2020,
doi = {10.3847/1538-3881/aba75b},
url = {https://dx.doi.org/10.3847/1538-3881/aba75b},
year = {2020},
month = {aug},
publisher = {The American Astronomical Society},
volume = {160},
number = {3},
pages = {123},
author = {Johnson, Samson A. and Penny, Matthew and Gaudi, B. Scott and Kerins, Eamonn and Rattenbury, Nicholas J. and Robin, Annie C. and Calchi Novati, Sebastiano and Henderson, Calen B.},
title = {Predictions of the Nancy Grace Roman Space Telescope Galactic Exoplanet Survey. II. Free-floating Planet Detection Rates*},
journal = {The Astronomical Journal}
}

@article{2012AA...538A.106R,
    title = {{Stellar populations in the Milky Way bulge region: towards solving the Galactic bulge and bar shapes using 2MASS data}},
    year = {2012},
    journal = {{\textbackslash}aap},
    author = {Robin, A.~C. and Marshall, D.~J. and Schultheis, M and Reyl{\'{e}}, C},
    month = {2},
    pages = {A106},
    volume = {538},
    doi = {10.1051/0004-6361/201116512},
    arxivId = {astro-ph.GA/1111.5744}
}

@ARTICLE{1978AnSta...6..461S,
       author = {{Schwarz}, Gideon},
        title = "{Estimating the Dimension of a Model}",
      journal = {Annals of Statistics},
         year = 1978,
        month = jul,
       volume = {6},
       number = {2},
        pages = {461-464},
       adsurl = {https://ui.adsabs.harvard.edu/abs/1978AnSta...6..461S}
}

@article{Penny2019,
    title = {{Predictions of the WFIRST Microlensing Survey. I. Bound Planet Detection Rates}},
    year = {2019},
    journal = {Astrophysical Journal, Supplement Series},
    author = {Penny, M.T. and Scott Gaudi, B. and Kerins, E. and Rattenbury, N.J. and Mao, S. and Robin, A.C. and Calchi Novati, S.},
    number = {1},
    volume = {241},
    doi = {10.3847/1538-4365/aafb69},
    issn = {00670049}
}

@article{Sandford_2017,
doi = {10.3847/1538-3881/aa94bf},
url = {https://doi.org/10.3847/1538-3881/aa94bf},
year = {2017},
month = {nov},
publisher = {The American Astronomical Society},
volume = {154},
number = {6},
pages = {228},
author = {Sandford, Emily and Kipping, David},
title = {Know the Planet, Know the Star: Precise Stellar Densities from Kepler Transit Light Curves},
journal = {The Astronomical Journal}
}

@ARTICLE{2007PASP..119..986B,
       author = {{Barnes}, Jason W.},
        title = "{Effects of Orbital Eccentricity on Extrasolar Planet Transit Detectability and Light Curves}",
      journal = {\pasp},
         year = 2007,
        month = sep,
       volume = {119},
       number = {859},
        pages = {986-993},
          doi = {10.1086/522039},
archivePrefix = {arXiv},
       eprint = {0708.0243},
 primaryClass = {astro-ph},
       adsurl = {https://ui.adsabs.harvard.edu/abs/2007PASP..119..986B}
}

@article{Twicken_2016,
doi = {10.3847/0004-6256/152/6/158},
url = {https://doi.org/10.3847/0004-6256/152/6/158},
year = {2016},
month = {nov},
publisher = {The American Astronomical Society},
volume = {152},
number = {6},
pages = {158},
author = {Twicken, Joseph D. and Jenkins, Jon M. and Seader, Shawn E. and Tenenbaum, Peter and Smith, Jeffrey C. and Brownston, Lee S. and Burke, Christopher J. and Catanzarite, Joseph H. and Clarke, Bruce D. and Cote, Miles T. and Girouard, Forrest R. and Klaus, Todd C. and Li, Jie and McCauliff, Sean D. and Morris, Robert L. and Wohler, Bill and Campbell, Jennifer R. and Uddin, Akm Kamal and Zamudio, Khadeejah A. and Sabale, Anima and Bryson, Steven T. and Caldwell, Douglas A. and Christiansen, Jessie L. and Coughlin, Jeffrey L. and Haas, Michael R. and Henze, Christopher E. and Sanderfer, Dwight T. and Thompson, Susan E.},
title = {DETECTION OF POTENTIAL TRANSIT SIGNALS IN 17 QUARTERS OF KEPLER DATA: RESULTS OF THE FINAL KEPLER MISSION TRANSITING PLANET SEARCH (DR25)},
journal = {The Astronomical Journal}
}

@article{10.1093/mnras/sty3463,
    author = {Zink, Jon K and Christiansen, Jessie L and Hansen, Bradley M S},
    title = {Accounting for incompleteness due to transit multiplicity in Kepler planet occurrence rates},
    journal = {Monthly Notices of the Royal Astronomical Society},
    volume = {483},
    number = {4},
    pages = {4479-4494},
    year = {2018},
    month = {12},
    issn = {0035-8711},
    doi = {10.1093/mnras/sty3463},
    url = {https://doi.org/10.1093/mnras/sty3463},
    eprint = {https://academic.oup.com/mnras/article-pdf/483/4/4479/27374488/sty3463.pdf},
}

@article{10.1093/mnras/sty1783,
    author = {Van Eylen, V and Agentoft, Camilla and Lundkvist, M S and Kjeldsen, H and Owen, J E and Fulton, B J and Petigura, E and Snellen, I},
    title = {An asteroseismic view of the radius valley: stripped cores, not born rocky},
    journal = {Monthly Notices of the Royal Astronomical Society},
    volume = {479},
    number = {4},
    pages = {4786-4795},
    year = {2018},
    month = {07},
    issn = {0035-8711},
    doi = {10.1093/mnras/sty1783},
    url = {https://doi.org/10.1093/mnras/sty1783},
    eprint = {https://academic.oup.com/mnras/article-pdf/479/4/4786/25193635/sty1783.pdf},
}

@ARTICLE{neptune_desert,
       author = {{Mazeh}, T. and {Holczer}, T. and {Faigler}, S.},
        title = "{Dearth of short-period Neptunian exoplanets: A desert in period-mass and period-radius planes}",
      journal = {\aap},
         year = 2016,
        month = may,
       volume = {589},
          eid = {A75},
        pages = {A75},
          doi = {10.1051/0004-6361/201528065},
archivePrefix = {arXiv},
       eprint = {1602.07843},
 primaryClass = {astro-ph.EP},
       adsurl = {https://ui.adsabs.harvard.edu/abs/2016A&A...589A..75M}
}

@ARTICLE{2025arXiv251026882W,
       author = {{Weldon}, Grant C. and {Hansen}, Bradley M.~S. and {Naoz}, Smadar},
        title = "{Saving Doomed Planets: Mass Loss and Angular Momentum Return Boost Hot Jupiter Survival Rates}",
      journal = {arXiv e-prints},
         year = 2025,
        month = oct,
          eid = {arXiv:2510.26882},
        pages = {arXiv:2510.26882},
          doi = {10.48550/arXiv.2510.26882},
archivePrefix = {arXiv},
       eprint = {2510.26882},
 primaryClass = {astro-ph.EP},
       adsurl = {https://ui.adsabs.harvard.edu/abs/2025arXiv251026882W}
}

@ARTICLE{2024A&A...689A.250C,
       author = {{Castro-Gonz{\'a}lez}, A. and {Bourrier}, V. and {Lillo-Box}, J. and {Delisle}, J.-B. and {Armstrong}, D.~J. and {Barrado}, D. and {Correia}, A.~C.~M.},
        title = "{Mapping the exo-Neptunian landscape: A ridge between the desert and savanna}",
      journal = {\aap},
         year = 2024,
        month = sep,
       volume = {689},
          eid = {A250},
        pages = {A250},
          doi = {10.1051/0004-6361/202450957},
archivePrefix = {arXiv},
       eprint = {2409.10517},
 primaryClass = {astro-ph.EP},
       adsurl = {https://ui.adsabs.harvard.edu/abs/2024A&A...689A.250C}
}

@article{10.1093/mnras/sty1760,
    author = {Owen, James E and Lai, Dong},
    title = {Photoevaporation and high-eccentricity migration created the sub-Jovian desert},
    journal = {Monthly Notices of the Royal Astronomical Society},
    volume = {479},
    number = {4},
    pages = {5012-5021},
    year = {2018},
    month = {07},
    issn = {0035-8711},
    doi = {10.1093/mnras/sty1760},
    url = {https://doi.org/10.1093/mnras/sty1760},
    eprint = {https://academic.oup.com/mnras/article-pdf/479/4/5012/25197597/sty1760.pdf},
}

@article{Morton_2016,
doi = {10.3847/0004-637X/822/2/86},
url = {https://doi.org/10.3847/0004-637X/822/2/86},
year = {2016},
month = {may},
publisher = {The American Astronomical Society},
volume = {822},
number = {2},
pages = {86},
author = {Morton, Timothy D. and Bryson, Stephen T. and Coughlin, Jeffrey L. and Rowe, Jason F. and Ravichandran, Ganesh and Petigura, Erik A. and Haas, Michael R. and Batalha, Natalie M.},
title = {FALSE POSITIVE PROBABILITIES FOR ALL KEPLER OBJECTS OF INTEREST: 1284 NEWLY VALIDATED PLANETS AND 428 LIKELY FALSE POSITIVES},
journal = {The Astrophysical Journal}
}

@ARTICLE{2018ApJS..235...38T,
       author = {{Thompson}, Susan E. and {Coughlin}, Jeffrey L. and {Hoffman}, Kelsey and {Mullally}, Fergal and {Christiansen}, Jessie L. and {Burke}, Christopher J. and {Bryson}, Steve and {Batalha}, Natalie and {Haas}, Michael R. and {Catanzarite}, Joseph and {Rowe}, Jason F. and {Barentsen}, Geert and {Caldwell}, Douglas A. and {Clarke}, Bruce D. and {Jenkins}, Jon M. and {Li}, Jie and {Latham}, David W. and {Lissauer}, Jack J. and {Mathur}, Savita and {Morris}, Robert L. and {Seader}, Shawn E. and {Smith}, Jeffrey C. and {Klaus}, Todd C. and {Twicken}, Joseph D. and {Van Cleve}, Jeffrey E. and {Wohler}, Bill and {Akeson}, Rachel and {Ciardi}, David R. and {Cochran}, William D. and {Henze}, Christopher E. and {Howell}, Steve B. and {Huber}, Daniel and {Pr{\v{s}}a}, Andrej and {Ram{\'\i}rez}, Solange V. and {Morton}, Timothy D. and {Barclay}, Thomas and {Campbell}, Jennifer R. and {Chaplin}, William J. and {Charbonneau}, David and {Christensen-Dalsgaard}, J{\o}rgen and {Dotson}, Jessie L. and {Doyle}, Laurance and {Dunham}, Edward W. and {Dupree}, Andrea K. and {Ford}, Eric B. and {Geary}, John C. and {Girouard}, Forrest R. and {Isaacson}, Howard and {Kjeldsen}, Hans and {Quintana}, Elisa V. and {Ragozzine}, Darin and {Shabram}, Megan and {Shporer}, Avi and {Silva Aguirre}, Victor and {Steffen}, Jason H. and {Still}, Martin and {Tenenbaum}, Peter and {Welsh}, William F. and {Wolfgang}, Angie and {Zamudio}, Khadeejah A. and {Koch}, David G. and {Borucki}, William J.},
        title = "{Planetary Candidates Observed by Kepler. VIII. A Fully Automated Catalog with Measured Completeness and Reliability Based on Data Release 25}",
      journal = {\apjs},
         year = 2018,
        month = apr,
       volume = {235},
       number = {2},
          eid = {38},
        pages = {38},
          doi = {10.3847/1538-4365/aab4f9},
archivePrefix = {arXiv},
       eprint = {1710.06758},
 primaryClass = {astro-ph.EP},
       adsurl = {https://ui.adsabs.harvard.edu/abs/2018ApJS..235...38T}
}

@ARTICLE{2013ARA&A..51..269D,
       author = {{Duch{\^e}ne}, Gaspard and {Kraus}, Adam},
        title = "{Stellar Multiplicity}",
      journal = {\araa},
         year = 2013,
        month = aug,
       volume = {51},
       number = {1},
        pages = {269-310},
          doi = {10.1146/annurev-astro-081710-102602},
archivePrefix = {arXiv},
       eprint = {1303.3028},
 primaryClass = {astro-ph.SR},
       adsurl = {https://ui.adsabs.harvard.edu/abs/2013ARA&A..51..269D}
}

@article{Raghavan_2010,
doi = {10.1088/0067-0049/190/1/1},
url = {https://doi.org/10.1088/0067-0049/190/1/1},
year = {2010},
month = {aug},
publisher = {The American Astronomical Society},
volume = {190},
number = {1},
pages = {1},
author = {Raghavan, Deepak and McAlister, Harold A. and Henry, Todd J. and Latham, David W. and Marcy, Geoffrey W. and Mason, Brian D. and Gies, Douglas R. and White, Russel J. and ten Brummelaar, Theo A.},
title = {A SURVEY OF STELLAR FAMILIES: MULTIPLICITY OF SOLAR-TYPE STARS},
journal = {The Astrophysical Journal Supplement Series}
}
%\end{thebibliography}

%%%%%%%%%%%%%%%%%%%%%%%%%%%%%%%%%%%%%%%%%%%%%%%%%%%%%%%%%%%%%%%
% Appendices must be placed after   \end{thebibliography}
% They will be placed automatically on a new page.
%%%%%%%%%%%%%%%%%%%%%%%%%%%%%%%%%%%%%%%%%%%%%%%%%%%%%%%%%%%%%%%
\begin{appendix}

\end{appendix}

\end{document}